\documentclass[aps, nofootinbib, twocolumn,groupedaddress,floatfix,nofootnotesinbib,preprintnumbers]{revtex4}
\usepackage{graphicx}
\usepackage{epsfig}
\usepackage{epstopdf}
\usepackage{rotating}
\usepackage[mathscr]{eucal}
\usepackage{amsmath,amssymb,bm,graphics}
\usepackage[pdftex,letterpaper=true]{hyperref}

\let\mathbf=\bm

\newif\ifarxiv
\arxivfalse

\begin{document}
\def\Nfour{\mathcal N\,{=}\,4}
\def\Ntwo{\mathcal N\,{=}\,2}
\def\Nc{N_{\rm c}}
\def\Nf{N_{\rm f}}
\def\x{\mathbf x}
\def\q{\mathbf q}
\def\f{\mathbf f}
\def\v{\mathbf v}
\def\C{\mathcal C}
\def\w{\omega}
\def\vs{v_{\rm s}}
\def\S{\mathcal S}
\def\half{{\textstyle \frac 12}}
\def\twothirds{{\textstyle \frac 23}}
\def\third{{\textstyle \frac 13}}
\def\t{\mathbf{t}}
\def\T{\mathcal {T}}
\def\O{\mathcal{O}}
\def\E{\mathcal{E}}
\def\p{\mathcal{P}}
\def\H{\mathcal{H}}
\def\uh{u_h}
\def\R{\ell}
\def\Ro{\chi}
\def\del{\nabla}
\def\eps{\hat \epsilon}
\def\nn{\nonumber}
\def\K{\mathcal K}
\def\inf{\epsilon}
\def\cs{c_{\rm s}}
\def\A{\mathcal{A}}
\def\e{{e}}
\def\r{{\xi}}
\def\x{{\mathbf x}}
\def\w{{w}}
\def\rr{{\xi}}
\def\uo{{u_*}}
\def\u{{\mathcal U}}
\def\G{\mathcal{G}}
\def\Deltax{\Delta x_{\rm max}}
\def\L{{\bm L}}

\title
{Shining a Gluon Beam Through Quark-Gluon Plasma}

\author{Paul~M.~Chesler\footnotemark}
\author{Ying-Yu Ho\footnotemark}
\author{Krishna~Rajagopal\footnotemark}

\affiliation
   {Department of Physics, Massachusetts Institute of Technology, Cambridge, MA 02139, USA}

\date{October 19, 2011}

\begin{abstract}
We compute the energy density radiated by a quark undergoing circular motion in strongly coupled 
$\mathcal N = 4$ supersymmetric Yang-Mills plasma.  
If it were in vacuum, this quark would radiate a beam of strongly coupled radiation whose angular distribution has been characterized and is very similar to that of synchrotron radiation produced by an electron in circular motion in electrodynamics.  Here, we watch this beam of gluons getting quenched by the strongly coupled plasma.  We find that 
a beam of gluons of momenta $\sim q \gg \pi T$  is attenuated rapidly, over a distance 
$\sim q^{1/3} (\pi T)^{-4/3}$ in a plasma with temperature $T$.   
As the beam propagates through the plasma at the speed of light, it
sheds  trailing sound waves with momenta $\lesssim \pi T$.
Presumably these sound waves would thermalize in the plasma if they were not hit 
soon after their production by the next pulse of gluons from the lighthouse-like rotating quark.    At larger and larger $q$, the trailing sound
wave becomes less and less 
prominent.
The outward going beam of gluon radiation itself 
shows no tendency to spread in angle or to shift toward larger wavelengths,
even as it is completely attenuated.  In this regard, the behavior of the beam of gluons that we analyze is reminiscent of the behavior of jets produced in heavy ion collisions at the LHC 
that lose a significant fraction of their energy without appreciable change in their angular distribution or their momentum distribution
as they plow through the strongly coupled quark-gluon plasma produced in these collisions.
\end{abstract}

\preprint{MIT-CTP-4309}


\maketitle
\iftrue
\def\thefootnote{\fnsymbol{footnote}}
\footnotetext[1]{Email: \tt pchesler@mit.edu}
\footnotetext[2]{Email: \tt yingyu@mit.edu}
\footnotetext[3]{Email: \tt krishna@mit.edu}
\def\thefootnote{\arabic{footnote}}
\fi

\parskip	2pt plus 1pt minus 1pt

\noindent \section{{Context}}  

Jet quenching is one of the most striking phenomena seen in heavy ion collisions at the Relativistic Heavy Ion Collider (RHIC) and at the Large Hadron Collider (LHC).   The basic picture behind a suite of observables measured in these collisions is that energetic partons produced via rare hard-scattering processes in heavy ion collisions lose a significant fraction of their energy as they and
the spray of partons they fragment into
plow through the strongly coupled plasma produced in the same collisions.   At the LHC, jet energies are high enough that the jets can be detected calorimetrically event-by-event, and the phenomenon of jet quenching is manifest in single events with, say, a jet with an energy greater than 200~GeV back-to-back with a jet with an energy less than 100~GeV~\cite{Aad:2010bu,Chatrchyan:2011sx,Angerami:2011is}.  (It is improbable that a pair of jets will be produced such that each travels the same distance through the plasma and loses the same amount of energy, so back-to-back pairs of jets with unbalanced energies are the norm.)    
At the LHC, the attenuated jets ({\it i.e.}~the lower energy jets in the unbalanced pairs) look remarkably like jets produced in vacuum, with angular distributions and momentum distributions (i.e. fragmentation functions) that are to date indistinguishable from those of jets with the same energy produced in proton-proton collisions~\cite{Chatrchyan:2011sx,Collaboration:2011yma,Collaboration:2011cs} or in peripheral heavy ion collisions~\cite{Steinberg:2011dj}.  The energy lost from the jets emerges instead as an excess of soft particles (momenta $\lesssim$ 1~GeV~\cite{Chatrchyan:2011sx})  at large angles ($>45^\circ$~\cite{Chatrchyan:2011sx}) relative to the jet direction.  

The observation of jets that have lost a large fraction of their energy was no surprise; it was surprising, however, to see the attenuated jets emerging without any visible softening of their fragmentation functions and without any visible broadening of their angular distribution.
In the limit of high parton energy $E$, the dominant energy loss process for an energetic parton plowing through quark-gluon plasma with temperature $T$ is gluon 
bremsstrahlung~\cite{Gyulassy:1993hr,Baier:1996sk,Zakharov:1997uu}, 
radiating gluons with energy $\omega$ and momentum transverse to the jet direction $k_\perp$ that satisfy $E\gg \omega \gg k_\perp \gg \pi T$~\cite{Baier:1996sk,Zakharov:1997uu,Wiedemann:2000za,Guo:2000nz,Gyulassy:2000er}. This set of approximations underlies all 
analytic calculations of radiative energy loss to date.
The (perhaps naive) expectation based upon these considerations is that at least some of the energy lost by the high energy parton should emerge as relatively hard particles (since $\omega \gg \gg \pi T$) near the jet direction (since $\omega \gg k_\perp$), resulting in a jet whose angular distribution has been broadened and whose fragmentation function has been softened.  
The data from the LHC have been stimulating more sophisticated implementations of 
these considerations~\cite{CasalderreySolana:2010eh,Qin:2010mn,Young:2011qx,
Zapp:2011ya,CasalderreySolana:2011rz,Qin:2011ng,CasalderreySolana:2011rq,ColemanSmith:2011rw,Young:2011va}, but it is also possible that the partons produced in LHC collisions are simply not energetic enough for this picture to apply at all --- since it is based upon the premise that the QCD coupling evaluated at the scale $k_\perp$ (which, recall, is $\ll \ll E$ but $\gg \pi T$) is weak even if the physics at scales $\sim \pi T$ is strongly coupled.  Given that this separation of scales may not be applicable, it behooves us to analyze jet quenching or models of jet quenching
in strongly coupled plasmas in contexts where reliable analyses are possible.  
Even if these analyses yield only qualitative insight, they
can be useful as benchmarks and as guides to how to think about the physics.  

The simplest strongly coupled plasma that we know of is that found at nonzero temperature in strongly coupled ${\cal N}=4$ supersymmetric Yang-Mills (SYM) theory in the large number of  colors ($N_c\rightarrow\infty$) limit, which is dual to the five-dimensional anti-deSitter 
black hole (AdS-BH)~\cite{Maldacena:1997re,Witten:1998qj,Witten:1998zw}.
(For a recent review of the by now many ways in which this theory and its many cousins have been 
used to gain insight into properties of strongly coupled plasma and phenomena in heavy ion collisions, see Ref.~\cite{CasalderreySolana:2011us}.)     Unfortunately, it is not possible to literally study jet quenching in ${\cal N}=4$ SYM theory because hard scattering in this theory does not produce jets~\cite{Hofman:2008ar}. Nevertheless, many authors have used the strongly coupled plasma  of ${\cal N}=4$ SYM theory to gain relevant insights, for example by studying the energy loss and momentum diffusion of a heavy quark plowing through the 
plasma~\cite{Gubser:2006bz,Herzog:2006gh,CasalderreySolana:2006rq,Gubser:2006nz,CasalderreySolana:2007qw} as well as the wake it produces~\cite{Friess:2006fk,Yarom:2007ni,Gubser:2007xz,Chesler:2007an,Gubser:2007ga,Gubser:2007ni,Chesler:2007sv,Noronha:2008un}, the stopping distance of an energetic light quark or gluon in this 
plasma~\cite{Gubser:2008as,Chesler:2008wd,Chesler:2008uy}, and the value of the jet quenching parameter~\cite{Liu:2006ug,D'Eramo:2010ak}, the property of the strongly coupled plasma that enters into the calculation of jet quenching if the scale-separation of the previous paragraph does turn out to be valid.   However,  none of these calculations yield insights into how an attenuated jet can emerge without any visible softening of its fragmentation function or 
spreading of its angular distribution. The one that comes closest is
the calculation of the stopping distance of 
a light quark~\cite{Gubser:2008as,Chesler:2008wd,Chesler:2008uy}.

Note that it does not make sense to interpret the data as saying that a single quark loses a lot of energy in the medium and then emerges in isolation and fragments into an ordinary-looking jet, since what would emerge is a nearly onshell quark, which would not fragment into a jet.
If we are to gain insight into jets in heavy ion collisions from a strongly coupled perspective, we should  imagine that the initial hard parton fragments quickly into a protojet of some sort and then ask how this protojet interacts with, and loses energy in, the strongly coupled plasma.
Although there is a sense in which the analyses of Refs.~\cite{Gubser:2008as,Chesler:2008wd,Chesler:2008uy} provide answers to this question, these results are
sensitive to details of the initial conditions, and we anyway do not expect strong coupling methods to describe the initial fragmentation of a hard parton into a protojet.   Instead, we shall assume that this initial stage proceeds conventionally, as described successfully 
in perturbative QCD, and then ask how the protojet interacts with, and loses energy in, the strongly coupled plasma.  
Recent work~\cite{Athanasiou:2010pv} offers a new way to gain a perspective on this question, since it provides a way to produce a beam of gluons whose angular distribution and wavelengths are well understood in vacuum.  In this paper, we shall shine such a beam of gluons through the strongly coupled plasma at nonzero temperature and watch this beam rapidly get attenuated --- without any apparent broadening of its angular distribution or lengthening of its wavelength!

\noindent \section{{Introduction}}  

The trick by which a beam of gluons can be produced in ${\cal N}=4$ SYM theory  is to consider a test quark undergoing circular motion
with radius $R_0$ and velocity $\beta$ (and hence angular velocity $\Omega\equiv \beta R_0$)
in the vacuum of this theory~\cite{Athanasiou:2010pv,Hubeny:2010bq}.  At both weak coupling (where the calculation is done conventionally) and strong coupling (where the calculation is done via gauge/gravity duality) the radiation that results is remarkably similar to the synchrotron radiation of classical electrodynamics, produced by an electron in circular motion~\cite{Athanasiou:2010pv}.  In particular, as the limit of ultrarelativistic motion is taken ($\gamma\to\infty$ where $\gamma\equiv 1/\sqrt{1-\beta^2}$) the lighthouse-like beam of radiation becomes more and more tightly collimated in angle (it is focused in a cone of angular extent $\sim 1/\gamma$) and is composed of gluons and scalars with shorter and shorter wavelengths (the pulse of gluons in the beam has a width $\sim R_0/\gamma^3$ in the radial direction  in which it is moving).
The emitted radiation was found to propagate outward  at the speed of light  forever
without broadening either in angle or in pulse width,
just as in classical electrodynamics~\cite{Athanasiou:2010pv,Hatta:2010dz,Hubeny:2010bq,Hatta:2011gh,Chernicoff:2011vn,Baier:2011dh,Chernicoff:2011xv}.  
At weak coupling, the slight differences in the angular distribution of the power radiated to infinity relative to that in classical electrodynamics can be attributed to the fact that scalars are radiated as well as gluons~\cite{Athanasiou:2010pv}. And, at strong coupling the angular distribution is identical to that at weak coupling~\cite{Athanasiou:2010pv,Hatta:2011gh,Baier:2011dh}.  

We shall compute the energy density radiated by a quark undergoing circular motion in strongly coupled ${\cal N}=4$ SYM theory at nonzero temperature, allowing us to watch what happens as the lighthouse beam of gluons and scalars is attenuated as it shines through the strongly coupled plasma.  We shall gain analytic understanding of the length scale over which the energy of the beam is attenuated by the plasma. And, by inspection of 
the energy density, which we obtain numerically, we shall see that as the beam 
is attenuated it does not broaden in angle or redden in wavelength.

The rate at which a quark undergoing circular motion
through the plasma of strongly 
coupled ${\cal N}=4$ SYM theory loses energy was studied 
previously in Ref.~\cite{Fadafan:2008bq}.  
In this analysis, two distinct regimes were found, depending on whether
\begin{equation}\label{XiDefn}
\Xi \equiv \frac{\Omega^2 \gamma^3}{(\pi T^2)}
\end{equation}
is $\gg 1$ or $\ll 1$.  For $\Xi \gg 1$, 
the energy loss rate is given by the generalized Larmor formula
\begin{equation}
\label{radiation}
\frac{dE}{dt} \Biggr|_{\rm rad}= \frac{\sqrt{\lambda}}{2 \pi}\, a^\mu a_\mu,
\end{equation}
where $a^\mu$ is the quark's proper acceleration and $\lambda\equiv g^2 N_c$ (with $g$ the gauge coupling) is the 't Hooft coupling which can be chosen at will since in this conformal theory it does not run and which we take as large since we wish to study strongly coupled plasma.
It was
shown many years ago by Mikhailov that the energy loss rate 
of a quark in circular motion in the vacuum of strongly coupled ${\cal N}=4$ SYM theory
is given by (\ref{radiation})~\cite{Mikhailov:2003er} and so in the strongly coupled
plasma at $T\neq 0$, in the $\Xi\gg 1$ 
regime we expect 
to see the radiation of
beam of synchrotron-like radiation as in vacuum~\cite{Athanasiou:2010pv}, and the subsequent attenuation of this beam.  When $\Xi\ll 1$, on the other hand, the energy loss rate is that due to the drag force exerted by the strongly coupled hydrodynamic fluid on a quark moving in a straight line with velocity $\beta$~\cite{Fadafan:2008bq}, namely~\cite{Gubser:2006bz,Herzog:2006gh}
\begin{equation}
\label{drag}
\frac{dE}{dt}\Biggr|_{\rm drag} = \frac{\sqrt{\lambda}}{2 \pi} \left(\pi T \right)^2 \beta^2 \gamma \ .
\end{equation}
Notice that the parameter $\Xi$ which governs which expression for the energy loss rate is 
valid is simply the ratio of the rates appearing  in Eqs.~(\ref{radiation}) and (\ref{drag}).  In this respect it is as if both hydrodynamic drag and Larmor radiation are in play with the larger of the two effects dominating the energy loss, but this simplified picture is not quantitatively correct because where $\Xi\sim 1$ the energy loss rate is less than the sum of Eqs.~(\ref{radiation}) 
and (\ref{drag})~\cite{Fadafan:2008bq}.
Although our principal interest is in the $\Xi > 1$ regime, where we can study the quenching of 
a beam of synchrotron gluons, it will prove instructive to look at $\Xi < 1$ and $\Xi\sim 1$ also as in these regimes the hydrodynamic response of the plasma --- i.e the production of sound waves --- is more readily apparent.

Unlike in vacuum, in the plasma at nonzero temperature
 the energy disturbance created by the rotating quark
can excite \textit{two} qualitatively distinct modes in the energy density; a sound mode which at long wavelengths travels at speed $c_s = 1/\sqrt{3}$, and a light-like mode
which propagates at the speed of light.  The relative amplitude of each mode depends
on the trajectory of the quark.  We find a correlation between the energy loss mechanism
and the relative amplitude of the light-like and sound modes that can be anticipated from 
the results of Ref.~\cite{Fadafan:2008bq}.  When $\Xi < 1$ the dominant modes 
that are excited
are sound waves.  When $\Xi > 1$ the dominant modes that are excited propagate at the speed of
light.  
Interestingly, by studying the $\Xi\sim 1$ regime we shall 
see that as the pulse of radiation moving at the
speed of light is attenuated in energy, it sheds a sound wave.

We shall use gauge/gravity duality to calculate the energy density that results when a test quark is moved through the strongly coupled plasma in a circle with $\Xi<1$ or $\Xi\sim 1$ or $\Xi > 1$. 
The gravitational dual of the undisturbed plasma is the (4+1)-dimensional AdS-Schwarzschild (AdS-BH) geometry.
Setting the AdS curvature radius $L = 1$, we choose coordinates such that the metric of the 
AdS-BH geometry is
\begin{equation}
\label{metric}
    ds^2 = \frac{1}{u^2}
    \left [-f(u) \, dv^2 + d \x^2 - 2\, dv\, du \right ] ,
\end{equation}
where $f(u) \equiv 1-(u/u_h)^4$.
These coordinates are generalized infalling Eddington-Finkelstein 
coordinates; lines of constant time $v$ represent infalling radial null
geodesics. The event horizon of the geometry is located at $u=u_h$,
with $T \equiv (\pi u_h)^{-1}$ the temperature of the equilibrium strongly coupled
${\cal N}=4$ SYM plasma that this metric describes. 
We shall do the calculation in
units where $\pi T = 1$, meaning $u_h=1$.
(However, in order to facilitate comparison to results at $T=0$ we shall report
results instead in units where $R_0=1$.)  The boundary of the AdS-BH metric, corresponding via
the holographic correspondence to the ultraviolet limit in the boundary ${\cal N}=4$ SYM quantum field theory, is at $u\rightarrow 0$.
We begin in Section III by reviewing the calculation of the shape of the string that hangs ``down'' from the rotating quark at the ultraviolet boundary of the AdS-BH toward its horizon, spiralling around and around infinitely many times just above the horizon.  This string profile was obtained in Ref.~\cite{Fadafan:2008bq}, but we rederive it in the coordinates that we shall use in Section~IV.  
In Section IV, we solve the bulk-to-boundary problem, finding the energy density in the boundary ${\cal N}=4$ SYM theory plasma that this spiralling string describes.  We describe our results in Section V and in Section VI we present analytic arguments that allow us to 
understand all of their qualitative features, and in Section VII we return to the context with which we begun, marvel at the qualitative resemblance between our results and jet quenching in heavy ion collisions at the LHC, and speculate on how this resemblance could be made more quantitative.

There are preliminary indications that the energy lost by  the lower energy jets (composed of longer wavelength gluons) studied at RHIC ends up in soft hadrons moving in directions correlated with the initial parton's path making the jets appear broadened in angle~\cite{Connors:2011zz,Caines:2011xp,Purschke:2011xx,Ohlson:2011xn,Caines:2011ew}, rather than far outside the jet 
cone as at the LHC~\cite{Chatrchyan:2011sx}.  We close by observing that our results 
illustrate a natural way for such a distinction between jet quenching at RHIC and the LHC to arise, since in our steady-state calculation 
we find that our beam of gluons excites less of a sound wave trailing behind the beam 
pulse --- aka soft particles going in roughly the jet direction --- when the beam is 
composed of shorter wavelength gluons than when it is composed of longer 
wavelength gluons.  We discuss this in Section VI.

\section{{String dynamics}} 

The dynamics of classical strings are governed by the Nambu-Goto action
$
%\begin{equation}
S_{\rm NG} = - T_0 \int d \tau \,d \sigma \sqrt{-g}\,,
$
%\label{SNG}
%\end{equation}
where $T_0 = \sqrt{\lambda}/{(2 \pi L^2)}$ is the string tension, $\sigma
$ and $\tau$ are the worldsheet coordinates of the string, and $g= {\rm det}\,
g_{ab}$ where $g_{ab}$ is
the induced worldsheet metric.  The string profile is determined by a
set of embedding
functions $X^M(\tau,\sigma)$ that specify where in the spacetime
described by the metric (\ref{metric}) the point $(\tau,\sigma)$ on the
string worldsheet is located. The induced world sheet metric is given
in terms of these functions by
%\begin{equation}
$g_{ab} = \partial_a X \cdot \partial_b X,$
%\end{equation}
where $a$ and $b$ each run over $(\tau,\sigma)$. 
For the determinant we obtain
\begin{equation}
-g= (\partial_\tau X \cdot \partial_\sigma X)^2 - (\partial_\tau X)^2
(\partial_\sigma X)^2\,.
\end{equation}

We choose worldsheet coordinates $\tau = v$ and $\sigma = u$.
As we are interested in quarks which rotate at constant frequency
$\Omega$ about the $\hat z$ axis, we parametrize the string embedding
functions
via
\begin{equation}
X^{M}(v,u) = (v,\bm r_s(v,u),u)\,,
\label{Xstring}
\end{equation}
where in spherical coordinates $\{r,\theta,\varphi\}$ the three-vector
$\bm r_s$ is given by
\begin{equation}
\bm r_s(v,u) \equiv \left (R(u), {\textstyle \frac{\pi}{2}},\varphi(u) +
\Omega v \right ).
\label{StringWorldsheetParametrization}
\end{equation}
With this parameterization, the Nambu-Goto action reads
\begin{equation}
\label{NB2}
S_{\rm NG} = - \frac{\sqrt{\lambda}}{{2 \pi}} \int dv \,du \, \mathfrak
L\,,
\end{equation}
where
\begin{equation}
\mathfrak L = \frac{\sqrt{1 -\Omega R^2 \left( \Omega R'^2 + 2 \varphi' \right ) + f \left( R'^2 + R^2 \varphi'^2 \right )}}
{u^2}\,.
\label{NGLagrangian}
\end{equation}

The equations of motion for $\varphi(u)$ and $R(u)$ follow from
extremizing the
Nambu-Goto action (\ref{NB2}).  One constant of the motion can be obtained
by noting
that the action is independent of $\varphi(u)$.  This implies that
\begin{equation}
\Pi \equiv  - \frac{\partial \mathfrak L}{\partial \varphi'}
\label{PiDef}
\end{equation}
is constant.   The minus sign on the right-hand side of (\ref{PiDef})  is there in order to make $\Pi$ positive for positive $\Omega$.
Eq.~(\ref{PiDef}) can be solved for $\varphi'(u)$ in terms of $R(u)$ and $R'(u)$.  There are two solutions for $\varphi'(u)$: one in which
$\varphi'(u)$ is regular near the horizon at $u = 1$ and one in which $\varphi'(u)$ diverges near the horizon.  In infalling Eddington-Finkelstein coordinates
the regular solution is aways the causal infalling solution and thus this is the one we choose.

The equation of motion for $R(u)$ is given by
\begin{equation}
\frac{\partial \mathfrak L}{\partial R} - \frac{\partial}{\partial u}
\frac{\partial \mathfrak L}{\partial R'}  = 0\,.
\end{equation}
Evaluating this expression and then
%Taking the above partial derivatives of $\mathcal L$ and then
eliminating $\varphi$ derivatives via Eq.~(\ref{PiDef}),
we obtain the following equation of motion:
\begin{align}
\label{Req}
R'' &+\frac{R \left(u + 2 R R' \right ) \left(1 + f R'^2 \right )}{u \left ( u^4 \Pi^2 - f R^2 \right)}
\\ \nonumber
 &+ \frac{u + u f R'^2 - 2 \left(1-f \right) RR' \left (1 + \Omega^2 R^2 R'^2 \right)}{u R \left (f - \Omega^2 R^2 \right )}  = 0.
\end{align}
Eq.~(\ref{Req}) is singular when
\begin{equation}
u^4 \Pi^2 - f R^2 = 0 \ \ {\rm or}\ \  f - \Omega^2 R^2 = 0.
\end{equation}
As discussed in Refs.~\cite{Fadafan:2008bq, Athanasiou:2010pv}, reality of the Nambu-Goto action (\ref{NB2}) implies that both singularities must coincide at a single value of $u$, which we denote 
$u = u_c$ upon defining 
\begin{equation}
\label{uc}
u_c \equiv \frac{1}{\sqrt{2}} \left [ \sqrt{4 {+}  \Pi^2 \Omega^2} -  \Pi \Omega \right ]^{1/2} 
\end{equation}
at which
\begin{equation}
\label{Rc}
R= R_c \equiv \sqrt{\frac{\Pi}{\Omega}} \, u_c \ .
\end{equation}
Furthermore, $g_{00}=0$ at the singular point $u=u_c$.  
Consequently, there is a worldsheet horizon 
at $u = u_c$, separating the upper part of the string $u<u_c$ which moves slower than the local velocity of light from the lower part of the string $u_c < u < u_h=1$ whose local velocity exceeds that
of light. The 
string equation (\ref{Req}) becomes first order at the worldsheet horizon,% 
\footnote
  {
  On can also show this directly 
  by solving Eq.~(\ref{Req}) via the Frobenius method, {\it i.e.} doing a Laurent 
  expansion about $u = u_c$.
  }
 meaning that the equation of motion (\ref{Req}) itself determines $R'(u)$ at $u=u_c$, independent of any features of the solution away from $(u_c,R_c)$.  Consequently, $\Pi$ and $\Omega$ determine not only $R(u_c)$ via Eqs.~(\ref{uc}) and~(\ref{Rc}) but also $R'(u_c)$, and then via the equation of motion (\ref{Req}) the entire solution $R(u)$~\cite{Fadafan:2008bq}.

For a given $\Pi$ and $\Omega$ we determine $R(u)$ numerically 
by solving Eq.~(\ref{Req}) using 
pseudospectral methods \cite{Boyd:2001}.  
We then solve Eq.~(\ref{PiDef}) for $\varphi(u)$, again using pseudospectral 
methods.

%We determined the string profile in Section III.A.   From this, we may now compute the $5d$ string stress
%tensor, which is what we need in order to determine the metric
%perturbation due to the string.  In general,
%\begin{equation}
%t^{MN}
%=
%-\frac{T_0}{\sqrt{-G}}\sqrt{-g}g^{ab}
%\partial_a X^M \partial_b X^N
%\delta^{3}(\bm r-\bm r_s)\,.
%\end{equation}

\section{{The bulk to boundary problem}}

 The presence of a string in the AdS-BH geometry, in our case the rotating string spiraling downward from the quark in circular motion at the boundary at $u=0$,
perturbs the geometry via Einstein's equations.  In the $\Nc \to \infty$ limit, the (4+1)-dimensional gravitational 
constant becomes parametrically small and the presence of the string acts as a small perturbation 
on the AdS-BH geometry.  To obtain leading order results in $\Nc$, we write the full 
(4+1)-dimensional 
metric as 
$G_{MN} = G_{MN}^{(0)} + \frac{L^2}{u^2} H_{MN}$, where $G_{MN}^{(0)} $ is the AdS-BH metric (\ref{metric}),
and linearize Einstein's equations in the perturbation $H_{MN}$.  This yields 
\begin{equation}
\label{linein}
\Delta^{AB}_{MN} H_{AB} = \kappa_5^2\, t_{MN},
\end{equation}
where $\Delta_{MN}^{AB}$ is a second order linear differential operator, $\kappa_5^2 = 4 \pi^2 /\Nc^2$
and 
\begin{equation}
t^{MN}
=
-\frac{T_0}{\sqrt{-G}}\,\sqrt{-g}\,g^{ab}
\partial_a X^M \partial_b X^N
\delta^{3}(\bm r-\bm r_s)
\end{equation}
is the (4+1)-dimensional string stress tensor.  The boundary value of the metric perturbation acts as a source for the ${\cal N}=4$ SYM 
stress tensor via the relation \cite{deHaro:2000xn}
\begin{equation}
T^{\mu \nu}(x) = 2 \frac{\delta S_{\rm grav}}{\delta \bar H_{\mu \nu}(x)} \bigg |_{\bar H_{\mu \nu} = 0},
\end{equation}
where $\bar H_{\mu \nu}(x)  \equiv \lim_{u \to 0} H_{MN}(x,u)$ and $S_{\rm grav}$ is the 
gravitational action.

The (3+1)-dimensional stress tensor $T^{\mu\nu}$ 
is symmetric, traceless and conserved and thus describes five independent 
degrees of freedom.  These degrees of freedom can  be conveniently 
packaged in five combinations of components of $T^{\mu \nu}$, each of which 
transforms with definite helicity under spatial rotations.
There are two helicity 2 components, two helicity 1 components and one helicity 0 
component of $T^{\mu \nu}$.
Similarly, $H_{MN}$ is a spin 2 field in (4+1) dimensions 
and thus contains 5 independent
gauge invariant helicity degrees of freedom, gauge invariant in
the sense that they are invariant under infinitesimal diffeomorphisms
\begin{equation}
\label{gaugetrans}
H_{MN} \to H_{MN} - D_M \xi_{N}- D_N \xi_{M},
\end{equation}
where $D_M$ is the covariant derivative with respect to the AdS-BH geometry and $\xi_M$ is an infinitesimal vector field.
Each gauge invariant helicity degree of freedom in $H_{MN}$ determines
the corresponding helicity component of $T^{\mu \nu}$ \cite{Chesler:2007sv, Chesler:2007an}.  Moreover, by rotational and gauge invariance, each 
gauge invariant helicity degree of freedom in $H_{MN}$ satisfies a decoupled equation of motion.
As we are only interested in the ${\cal N}=4$ SYM energy density, 
we focus below on helicity 0 gauge invariants.

To determine a helicity 0 gauge invariant and its equation of motion, it is convenient to 
decompose $H_{MN}$ in terms of a complete set of functions $\chi_q(x)$ of the $4d$ coordinates $x^\mu$.
We choose 
\begin{equation}
\chi_q(x) = e^{-i \omega v}\psi_q(\bm x), \ 
\end{equation}
where $\psi_q$ are eigenfunctions of the spatial Laplacian $-\del^2 \psi_q = q^2 \psi_q$.  In what follows, 
we shall initially not choose a specific basis of eigenfunctions $\psi_q$.
% and make no assumptions about the string stress tensor.
%We shall later simplify things by using the fact that the motion we are interested in is periodic in time.
% (\textit{e.g.} periodicity in time).
We define 
\begin{align}
\H_{MN} &\equiv  \int d^4 x\, \chi_q^*\, H_{MN}, \ \
\H_{qq} \equiv  -\frac{1}{q^2} \int d^4 x\, \chi_q^*\, \del_i \del_j H _{ij}, 
\\
\H_{q5} &\equiv  \frac{1}{i q} \int d^4 x\, \chi_q^*\, \del_i H_{i5}, \ \ 
\H_{0q} \equiv  \frac{1}{i q} \int d^4 x \,\chi_q^*\, \del_i H_{0i}, 
\end{align} 
and define similar expressions for $t_{MN}$ with the replacements
$H_{MN} \to t_{MN}$ and $\H_{MN} \to \t_{MN}$.
(If the $\psi_q$ were taken to be plane waves then $\H_{MN}$ would be 
the Fourier transform of $H_{MN}$.)
Note that the $q$ subscripts are labels, not values of the indices $M$ or $N$, while the $i$ and $j$ are spatial indices corresponding to values of $M$ or $N$ given by 1, 2 or 3. When $i$ or $j$ is repeated, this indicates summation.
With these definitions in hand, the helicity 0 field that we shall use is given by
\begin{align}
\nonumber
Z &\equiv \frac{4 q f^2 }{\omega}\partial_u \frac{\H_{0q}}{f^2} +\left (2 u q^2 {-}3 f'\right) \H_{qq} +\left (f' {-} 2 u q^2 \right) \H_{ii}   \\ \label{Zdef}
&  - \frac{4 q^2}{i \omega}\H_{00} +  4 i q f \H_{q5} + \frac{4 q^2 f}{i \omega} \H_{05} 
+\frac{8 \kappa_5^2 }{i \omega} \left ( \t_{00}-f \t_{05}   \right ),
\end{align}
As can be easily verified, $Z$ is invariant under the infinitesimal diffeomorphisms (\ref{gaugetrans}).
%This gauge invariant is the same one used in [[our old paper]] but expressed in terms of Eddington-Finkelstein time.
From the linearized Einstein equations (\ref{linein}) it is straightforward but tedious to show that  
$Z$ satisfies 
\begin{equation}
\label{Zeq}
\mathcal L Z = \kappa_5^2 \,S,
\end{equation}
where the linear operator $\mathcal L$ is given by 
\begin{align}
\nonumber
\mathcal L &\equiv f \frac{d^2}{du^2} + \left [ \frac{q^2 \left(u^4-5\right)+6 u^2 \left(5 u^4-9\right)}{u \left(q^2+6 u^2\right)}+2 i \omega \right ] \frac{d}{du}
\\ \label{lindef}
&+\frac{q^2 \left(5 u^4{-}5 i u \omega {+}9\right)-18 u^2 \left(3 u^4{+}3 i u \omega {-}7\right) -q^4u^2 }{u^2 \left(q^2+6 u^2\right)},
\end{align}
and the source $S$ is given by
\begin{align}
\nonumber 
S &\equiv 8\partial_u \t_{00}  - \frac{16 \left ( q^2 {+} 18 u^2 \right)}{u \left (q^2 {+} 6 u^2 \right)} \t_{00}
+ \frac{4 u \left (q^2 {+} 6 u^2\right )}{3}\left(  \t_{ii} - 2  \t_{qq} \right ) 
\\
&+ 8 i q (\t_{0q} - f \t_{q5} )
 - \frac{8 u  q^2 f }{3} \t_{55} + \frac{8}{3} \left (2 u q^2 {+} 3 i \omega \right ) \t_{05}.
 \label{SourceDefn}
\end{align}

For strings which end at the boundary $u = 0$, the source has the expansion
\begin{equation}
S = S_{(0)} + S_{(1)} u + {\cal O}(u^2),
\end{equation}
where 
\begin{eqnarray}
S_{(0)} &=& -8 \lim_{u \to 0}  u^2 \partial_u \left (\frac{\t_{00}}{u^2} \right ) = -8 \lim_{u \to 0} \partial_u \t_{00}\nonumber\\
 &=& -\frac{\sqrt{\lambda}}{2 \pi} \frac{16}{%L^3 
 \sqrt{1 - R_0^2 \Omega^2}} \chi_q^*(R_0)\ ,
\end{eqnarray}
a result that follows from Eq.~(\ref{SourceDefn}) upon noting that
$\t_{00}$ and indeed all the components of $\t$ in Eq.~(\ref{SourceDefn}) are  proportional to 
$u$ for small $u$.
Solving Eq.~(\ref{Zeq}) with a series expansion near $u=0$ and demanding
$\lim_{u \to 0} H_{MN} = 0$, we find 
\begin{equation}
\label{Zexp}
Z = Z_{(2)} u^2 + Z_{(3)} u^3 + \dots.
\end{equation}
To lowest nontrivial order in $u$, Eq.~(\ref{Zeq}) becomes just
$Z_{(2)} = S_{(0)}$, but we shall also need $Z_{(3)}$ because  the change in the ${\cal N}=4$ SYM energy density due to the presence of the moving quark (i.e. the total energy density minus that of the undisturbed plasma) is given by \cite{Athanasiou:2010pv}
\begin{equation}
\label{bndenergy}
\Delta \mathcal E = -\frac{1}{8 \kappa_5^2} \left (Z _{(3)} + i \omega Z_{(2)} \right ).
\end{equation}
This is as far as we can proceed without specifying a choice of the basis functions $\psi_q$.

%To compute the energy density for a given string profile, we exploit the cylindrical symmetry of the string and 
We choose to compute the boundary theory energy density using
the basis functions 
\begin{equation}
\psi_q(\bm x) =  j_{\ell}(q r)\, Y_{\ell m}(\theta,\phi),
\end{equation}
where $j_\ell$ are spherical Bessel functions and $Y_{\ell m}$ are spherical harmonics.
Our calculation is much simpler than the calculation for a quark moving along an arbitrary trajectory because, as in Ref.~\cite{Athanasiou:2010pv}, 
the quark has been moving in a circle at a constant angular velocity $\Omega$ for all time, meaning that everything in the problem (the 4+1 dimensional string stress tensor and the 3+1 dimensional boundary theory energy density) is rotating at this constant angular velocity and depends on the azimuthal angle $\phi$ and the time $v$ only in the combination $\phi-\Omega v$.  (At the $u=0$ boundary, Eddington-Finkelstein time $v$ is just boundary theory time $t$.)   
%Moreover, as the $5d$ string stress tensor only depends on the combination $\Omega v - \phi$, 
It follows that 
the energy density only has support when $\omega = m \Omega$.   Hence, the Fourier transform in time is a discrete Fourier transform.  Furthermore, 
specifying $m$ specifies $\omega$.  

For each $\{q, \ell, m\}$ we solve Eq.~(\ref{Zeq}) using pseudospectral methods in which the $u$-dependence of all functions is decomposed in terms of Chebyshev polynomials~\cite{Boyd:2001}.  The requisite two boundary conditions are specified at 
$u = 0$ and $u = 1$, as we now describe.  At $u = 0$, there are two linearly independent solutions.  The one we want is regular at $u=0$ and has the expansion (\ref{Zexp}), while the one that must be taken to vanish
 is proportional to $u^3 \log u$ at small $u$.  Imposing a boundary condition that 
 selects the regular solution is made easier 
  by first
 writing 
 \begin{equation}
 Z=u^2 S_{(0)} + u^3 X
 \end{equation}
 and turning the differential equation 
 for $Z$ into a differential equation for $X$.  In this differential equation, $X''$ (by $'$ we mean $\partial_u$) arises only in the term $ufX''$, meaning that at $u=0$ the differential equation for $X$ 
 involves only $X$ and $X'$, as long as $X''$ is finite.  So, the differential equation itself specifies its own boundary condition (as a relation between $X$ and $X'$) at $u=0$.  Satisfying this boundary condition automatically yields a regular solution at $u=0$, since if $X$ is finite the boundary condition makes $X'$ finite, and if $X'$ is finite the boundary condition makes $X''$ finite, and so on.
(The undesirable solution has $X$, $X'$ and $X''$ all divergent at $u=0$.) 
% the boundary condition is simply that of the 
%series expansion (\ref{Zexp}).  
At the AdS-BH horizon $u=1$, there are again two linearly independent solutions and we must specify a boundary condition that selects only the infalling mode.  Because we are using ingoing Eddington-Finkelstein coordinates, this boundary condition is equivalent to requiring that $Z$ (or $X$) and all of its derivatives are finite at $u=1$.  Again, as long as $X''$ is finite at $u=1$ the term $ufX''$ vanishes there, this time because $f=0$ at $u=1$, and the differential equation for $X$ turns into a boundary condition relating $X$ and $X'$ which, when satisfied, yields a regular solution at $u=1$.
(In ingoing Eddington-Finkelstein coordinates, the outgoing mode has a divergent phase
at $u=1$.)
%At the AdS-BH horizon at $u = 1$, we impose the boundary 
%condition of regularity, which is tantamount to infalling boundary conditions.  With this condition,
%the equations of motion (\ref{Zeq}) become first order at the horizon.  

Once the solution
$Z(q,\ell,m,u)$ is determined, we extract $Z_{(3)}$ (which is just $X$ evaluated at $u=0$)
and compute the Fourier components of the energy density $\Delta \mathcal E(q,\ell,m)$ via (\ref{bndenergy}).  The real-space
energy density is then computing by evaluating%
\begin{equation}
\Delta \mathcal E(r,\theta,\phi) = \frac{2}{\pi} \sum_{\ell m} \int q^2 dq \ \Delta \mathcal E(q,\ell,m) \,j_{\ell}(q r) \,Y_{\ell m}(\theta,\phi)
\label{FourierTransform}
\end{equation}
numerically.
This is the change in the energy density at time $t=0$ 
relative to the energy density of the unperturbed plasma
due to the presence of the rotating quark. The time-dependent energy density $\Delta \mathcal E(r,\theta,\phi,t)$ is 
obtained simply by replacing $\phi$ by $\phi-\Omega t$.  In order to obtain the results that we 
illustrate in the next Section, we typically used between $10^3$ and $10^4$ values of $q$ between $0$ and $200\, \pi T$ and at least 40 values of $\ell$, with smooth window functions cutting the Fourier transform (\ref{FourierTransform}) off at large $q$ and $\ell$.

One further complication remains, before we turn to our results.  The differential operator ${\mathcal L}$ and the source $S$ in (\ref{lindef}) and (\ref{SourceDefn}) are badly behaved at $u=0$ for small $q$.  There is no problem of principle, but solving  the equation (\ref{Zeq}) numerically becomes intractable.  Because the energy density has support only where $q \gtrsim \omega = m\Omega$, this difficulty only arises for the $m=0$ modes.  We therefore use the procedure that we have described only for the $m\neq 0$ components of the energy
density  $\Delta \mathcal E(q,\ell,m)$.  We calculate the $m=0$ components using a different gauge invariant helicity 0 field $Z_0$ that we define and describe in Appendix A.

\section{Results}

\begin{figure*}[ht]
\includegraphics[scale=0.49]{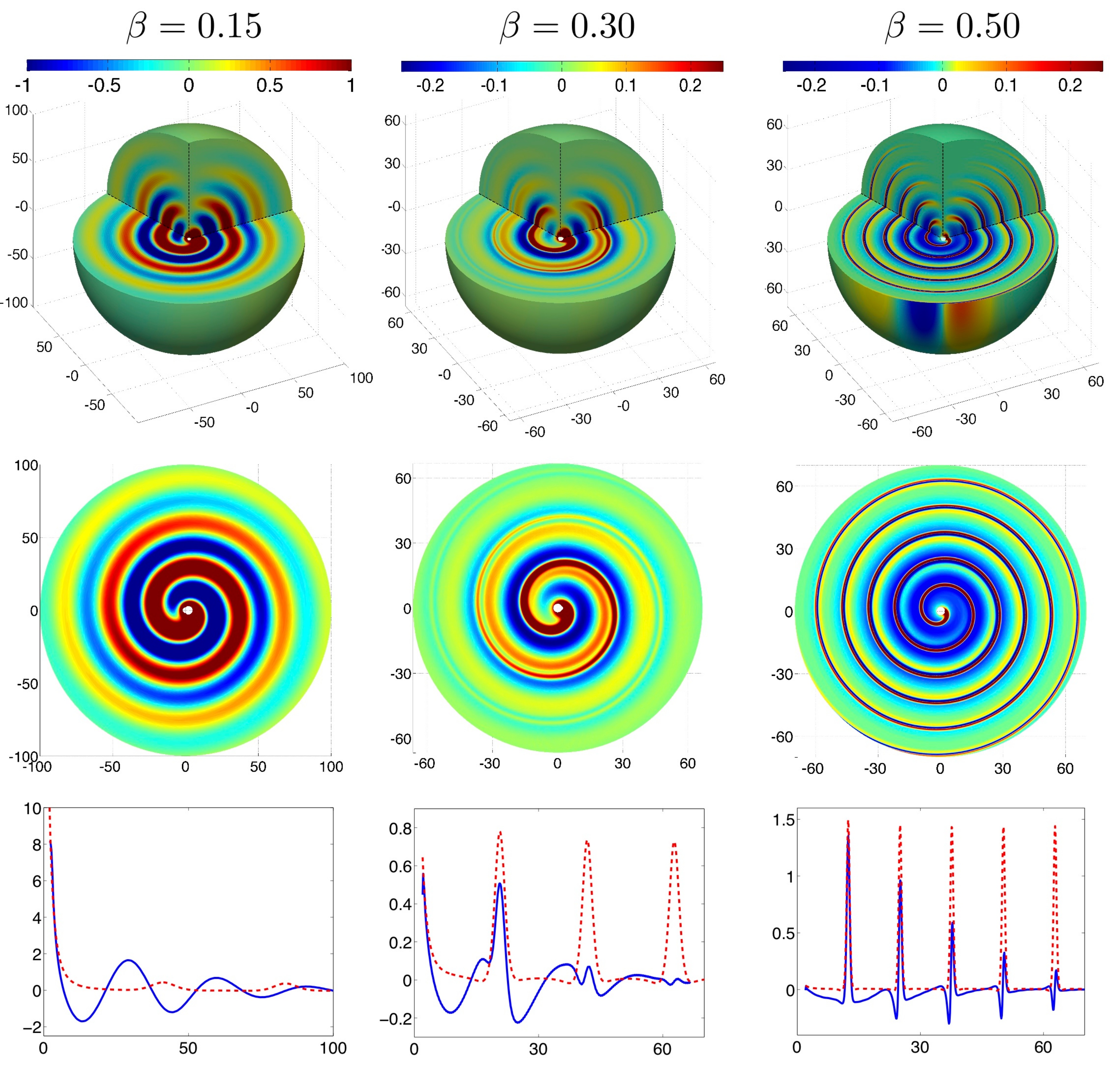}
\caption
    {\label{blingbling}
    Plots illustrating the energy density of strongly coupled ${\cal N}=4$ SYM plasma in which a
    test quark is rotating on a circle with radius $R_0$ with angular velocity $\Omega=\beta /R_0$ for $\beta=0.15$ (left column), $\beta=0.3$ (middle column) and $\beta=0.5$ (right column), corresponding to $\Xi=1.0$, 4.6 and 17.1.  In all plots, the temperature of the plasma is given by
    $\pi T = 0.15/R_0$ and the 
    units are chosen such that $R_0=1$.
       Top: cutaway plots of $r^2 \Delta \mathcal E/P$ where $P$ is the power radiated by the quark.  The cutaways coincide with the planes 
   $\theta = \pi/2$, $\phi = 0$ and $\phi =  7 \pi / 5$.  Middle: plots of $r^2 \Delta \mathcal E/P$
   on the equatorial plane $\theta=\pi/2$ (i.e. $z = 0$).  Bottom: blue curves are plots of $r^2 \Delta \mathcal E/P$ at $\theta=\pi/2$ and $\phi = \pi/2$.  
     % In all plots, units are chosen such that the radius of the quark's trajectory is $R_0=1$, and the temperature of the plasma is given by $\pi T = 0.15/R_0$.
   The quark's trajectory lies in the equatorial plane $\theta=\pi/2$ and the quark is rotating 
   counter-clockwise.  
   The red dashed curves in the bottom plots show $r^2 \mathcal E/P$ for 
   the strongly coupled synchrotron radiation emitted by a quark in circular motion {\it in vacuum}~{\protect\cite{Athanasiou:2010pv}}, pulses of radiation that propagate outward to $r\rightarrow \infty$ at the speed of light without spreading.
    %At both velocities
   %the energy radiated by the quark is concentrated along a spiral structure
   %which propagates radially outwards at the speed of light.  
   %The spiral is 
   %ocalized about $\theta = \pi/2$ with a characteristic width $\delta \theta \sim 1/\gamma$.
   %As $v \to 1$ the radial thickness $\Delta$ 
   %of the spirals rapidly decreases like $\Delta \sim 1/\gamma^3$.  
   }
\end{figure*}

Fig.~\ref{blingbling} shows three different plots of $r^2 \Delta \mathcal E/P$ for quarks in circular motion with each of three different velocities:
$\beta = 0.15$, $\beta = 0.3$ and $\beta = 0.5$.
Here, $P\equiv dE/dt$ is the energy lost by the circulating quark (and hence dumped into
the plasma) per unit time.
The radius of the quark's trajectory in all plots is $R_0 = 1$ and the temperature of the plasma is given by $\pi T = 0.15/R_0$.  This means 
that $\Xi$ defined in (\ref{XiDefn}) is given by 1.0, 4.6 and 17.1 in the left, middle and right columns respectively.
$P$ is given by (\ref{radiation}) when $\Xi\gg 1$ or by (\ref{drag}) when $\Xi\ll 1$. $P$ has been calculated in Ref.~\cite{Fadafan:2008bq} for any $\Xi$ and is related to the constant $\Pi$ defined in (\ref{PiDef}) by
\begin{equation}
P = \frac{\sqrt{\lambda}}{2 \pi} \Pi \Omega = \frac{\sqrt{\lambda}}{2 \pi} \frac{\Pi\beta}{ R_0}\ .
\end{equation}
In the left, middle and right columns of Fig.~\ref{blingbling}, $\Pi$ is given by
$5.2\times 10^{-3}/R_0$, $3.0 \times 10^{-2}/R_0$ and $0.22/R_0$ respectively.
%, which correspond to $(\Omega =0.15/R_0, \Pi =5.2\times 10^{-3}/R_0)$, $(\Omega =0.3/R_0, \Pi =3.0\times 10^{-2}/R_0)$ and $(\Omega =0.5/R_0, \Pi =0.22/R_0)$ respectively.  
At the time shown, the quark is located at $x = R_0$, $y = 0$ and the quark is rotating 
counter-clockwise  in the plane $z = 0$. 
The three plots in the top row are cutaway plots with the cutaways coinciding with the planes 
$z = 0$, $\phi = 0$ and $\phi =  7 \pi / 5$.  The three plots in the middle row show the energy density on the plane $z = 0$
and the bottom three plots give the energy density at $z = 0$, $\phi = \pi/2$, namely a slice through the middle-row plot along one radial line.  
For reference, the red dashed curves in these bottom plots 
show $r^2{\mathcal E}$ for the strongly coupled synchrotron radiation that a quark moving along the  same circular trajectory would emit in vacuum~\cite{Athanasiou:2010pv}.  
In each of the bottom plots, we use the same $P$ to normalize the red curve as for the blue curve.
All nine panels in Fig.~\ref{blingbling}
show the energy density at one instant of time, but the time-dependence is easily restored by replacing the azimuthal angle $\phi$ by $\phi-\Omega t$, where $\Omega=\beta /R_0$ is the angular velocity. 
As a function of increasing time, the entire patterns in the upper and middle rows rotate with angular velocity $\Omega$, as the spirals of radiation move outwards.  As a function of increasing time, the patterns in the lower rows move outwards, repeating themselves after a time $2\pi/\Omega$. 

As is evident from Fig.~\ref{blingbling}, as the quark accelerates along its circular trajectory, energy is radiated outwards in a spiral pattern which is attenuated as the radiation propagates outwards through
the plasma to increasing $r$.  
However, the qualitative features of the spiral patterns differ greatly at the three different quark velocities that we have chosen.  For $\beta = 0.15$ the spiral arms are very broad in 
$r$, as broad as their separation, and the spiral pattern propagates outwards at the speed of sound, while being attenuated with increasing $r$.  (Second-order hydrodynamics for a conformal fluid 
with a gravity dual like ${\cal N}=4$ SYM theory
predicts a sound velocity $1/\sqrt{3} + 0.116 \,q^2/(\pi T)^2 + \ldots$~\cite{Baier:2007ix} for sound waves with wave vector $q$. The sound waves in the left column of Fig.~\ref{blingbling} have $q\sim 1.3\, \pi T$ and are moving outward with a velocity of 0.73.  We shall return to the comparison to second-order hydrodynamics in Section VI.)  The red curve in the lower-left panel shows the energy density of the synchrotron radiation that this quark would have emitted if it were in vacuum, and we see that there is no sign of this in our results.  So, at this $\beta$, corresponding to $\Xi=1.0$, the rotating quark is emitting sound waves.

\begin{figure}[t]
\includegraphics[scale=0.35]{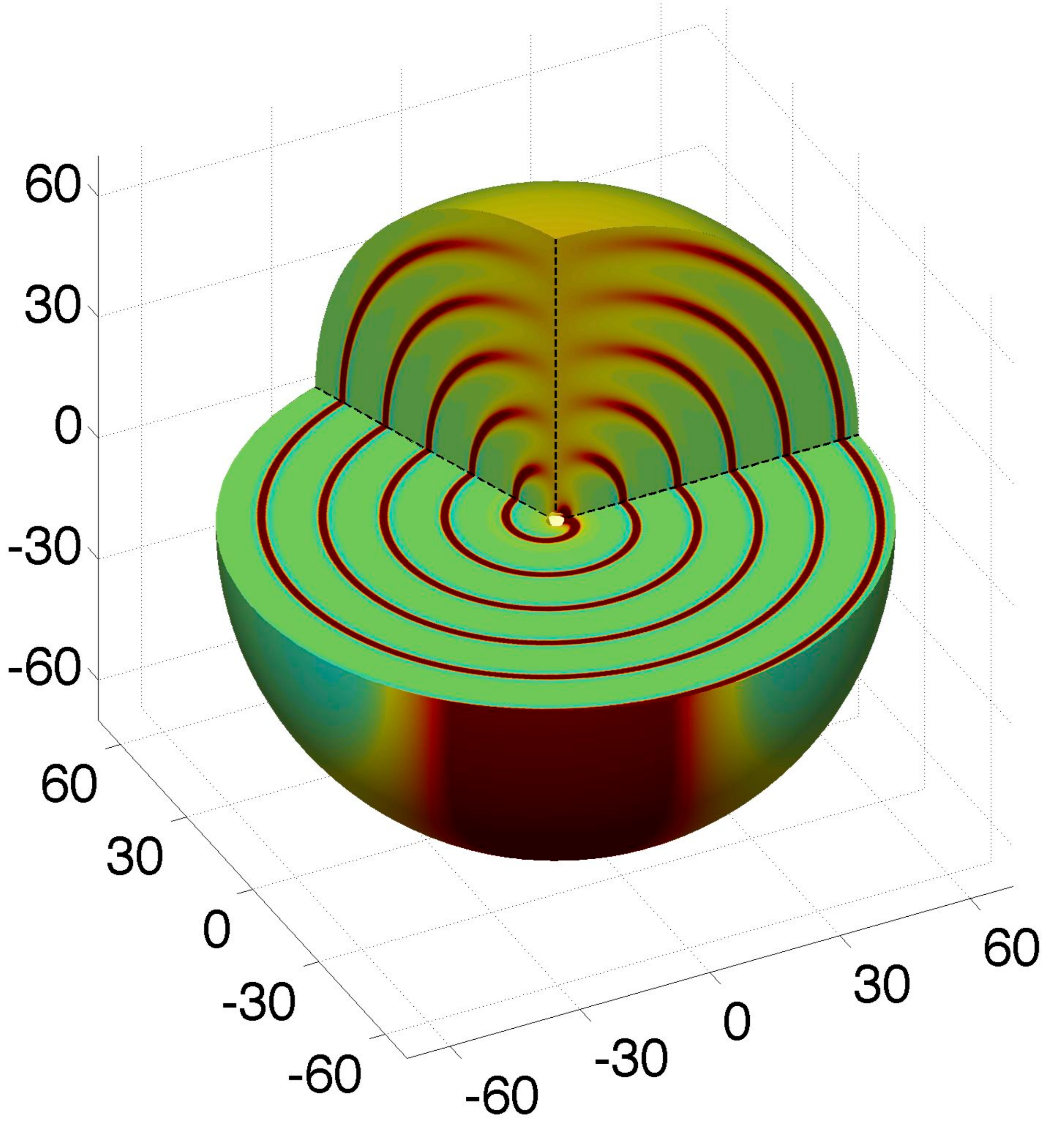}
\caption
    {\label{ZeroTRad}
    Energy density $r^2 \mathcal E/P$ of the strongly coupled synchrotron radiation emitted by a quark in circular motion with $R_0=1$ and $\beta=0.5$, exactly as in the top-right panel of Fig.~{\protect\ref{blingbling}}, but this time in vacuum, at $T=0$, calculated as in Ref.~{\protect\cite{Athanasiou:2010pv}}.  The color-scale is the same as in the top-right panel of Fig.~{\protect\ref{blingbling}}, and  the red curve in the bottom-right panel of Fig.~{\protect\ref{blingbling}} is the profile along one radial line through this figure.
             }
\end{figure}

The results in the right column of 
Fig.~\ref{blingbling}, for $\beta=0.5$, are strikingly different.  The spiral arms are very narrow in $r$, much narrower than their separation, and they propagate outwards at the speed of light, as can be seen immediately in the bottom-right panel  by comparing our results, in blue, to the energy density of the synchrotron radiation that this quark would have emitted if it were in vacuum.  We see that at this $\beta$, corresponding to $\Xi=17.1$, the rotating quark is emitting strongly coupled synchrotron radiation, as in vacuum~\cite{Athanasiou:2010pv}, and we see  that the radiation is being attenuated as it propagates outward in $r$, through the strongly coupled plasma.  Remarkably, even as the outgoing pulses of energy are very significantly attenuated by the medium
we see no sign of their broadening in either the $\theta$ or the $\phi$ or the $r$ directions.  Looking at the vertical sections in the upper-right panel, we see that if anything the spread of the beam of radiation in $\theta$ is becoming less as it propagates and gets attenuated. This conclusion is further strengthened by comparing the upper-right panel of Fig.~\ref{blingbling} to the analogous results for a quark in circular motion in vacuum shown in Fig.~\ref{ZeroTRad}.   It is certainly clear that the presence of the medium does {\it not} result in the spreading of energy away from the center of the beam at the equator out toward large polar angles.  Just the opposite, in fact: at large polar angles the beam gets attenuated more rapidly than near $\theta=\pi/2$.  We shall return to this in Section VI.
Broadening in either the $\phi$ or the $r$ directions would be manifest as widening of the pulses in the bottom-right panel, and this is certainly not seen.  In fact, we have extended the plot in the bottom-right panel out to larger $r$, for several more turns of the spiral, and we continue to see rapid attenuation with no visible broadening.  

We turn our attention now to the center column of Fig.~\ref{blingbling}.  Here, with a rotation velocity of $\beta=0.3$ corresponding to $\Xi=4.6$, we clearly see both synchrotron radiation and sound waves.  The synchrotron radiation is most easily identified with reference to the results for a quark with this rotation velocity in vacuum, shown in the red curve in the bottom-center panel. In our results with $T\neq 0$, we see the emission of a pulse of synchrotron radiation whose amplitude is very rapidly attenuated, much more rapidly than in the right column.  In part guided by our inspection of the results at large $\Xi$ in the right column, we see that as the pulse of synchrotron radiation is attenuated, it too does not broaden.  What we see here that is not so easily seen in the right column is that as the pulse of synchrotron radiation is attenuated it ``sheds'' a sound wave, leaving behind it a broad wave, reminiscent of the sound waves in the left column.   Behind each pulse of synchrotron radiation we see the ``compression half'' of a sound wave, and behind that a deeper rarefaction, and then the next pulse of synchrotron radiation arrives.  Once seen in the middle column, this phenomenon can perhaps also be discerned to a much lesser degree
 in the right column, with the each pulse of synchrotron radiation trailed first by a slightly yellow region of compression and then by a more blue region of rarefaction.  It is not really clear in the right column whether these can be called sound waves, both because of their smaller amplitude and because the next pulse of synchrotron radiation overwhelms them sooner than in the middle column.  In the middle column, though, the interpretation is clear:  the beam of synchrotron gluons is %losing energy by 
 exciting sound waves in the plasma.  In the right column, it is clear that any sound waves, if present, are smaller in amplitude than in the middle column even though the
 pulses of radiation carry more energy in the right column.

\begin{figure}[t]
\vspace{-0.28in}
\includegraphics[scale=0.517]{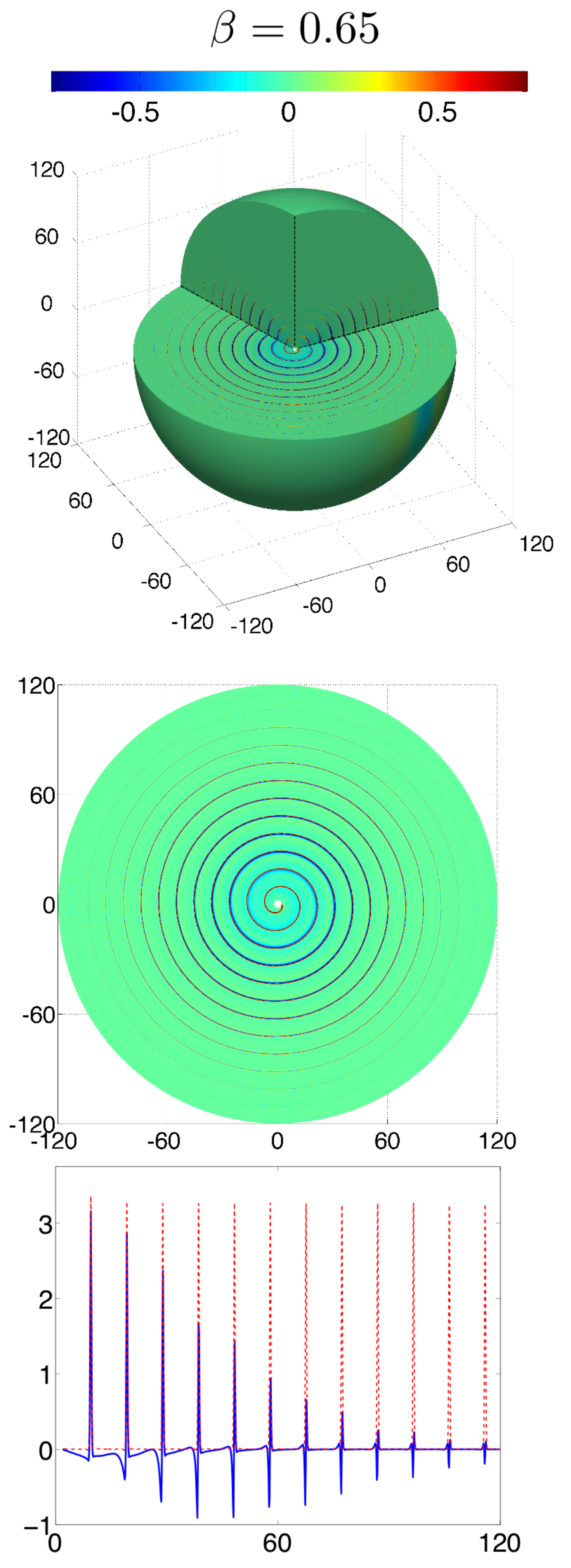}
\vspace{-.05in}
\caption
    {\label{Beta065blingbling}
      Energy density $r^2 \Delta \mathcal E/P$ of strongly coupled ${\cal N}=4$ SYM plasma in which a
    quark is rotating on a circle with $R_0\pi T=0.15$ and with velocity $\beta=0.65$, meaning $\Xi=42.8$. Plots as in one column of Fig.~{\protect\ref{blingbling}},   
    %As in that figure, $\pi T R_0=0.15$ and the units in the plot are chosen
    %such that $R_0=1$. 
    extended to $r=120$ to show the quenching of the
    beam of strongly coupled radiation.
             }
\vspace{-0.195in}
\end{figure}

Fig.~\ref{blingbling} demonstrates clearly that at small $\Xi$ the rotating quark emits only sound waves while at large $\Xi$ it emits strongly coupled synchrotron radiation as in vacuum, with that beam of gluons subsequently being quenched by the plasma.
%with a fraction of the energy lost by the beam that decreases with increasing $\Xi$ appearing
%initially as sound waves trailing behind the pulse of radiation.
 The crossover between these two regimes lies between $\Xi=1$ and $\Xi=5$, consistent with expectations based upon the results of Ref.~\cite{Fadafan:2008bq}.  We have also confirmed that this crossover occurs in the same range  of $\Xi$ for a quark in circular motion with $R_0 \pi T$ twice as large as in Fig.~\ref{blingbling}, meaning that this range of $\Xi$ occurs at larger $\beta$.  For $R_0 \pi T=0.15$ as in Fig.~\ref{blingbling}, we have looked at a quark moving 
with $\beta=0.10$, meaning $\Xi=0.45$ and confirmed that we see only sound waves, with longer wavelengths than in the left column of Fig.~\ref{blingbling} and hence with a velocity that is closer to $1/\sqrt{3}$, the $q\rightarrow 0$ velocity of sound.   We have also done the calculation at $\beta=0.65$, corresponding 
to $\Xi=42.8$, with results  shown in Fig.~\ref{Beta065blingbling} which we have extended out to larger $r$ to show the beam of synchrotron radiation getting almost completely attenuated.  The comparison in the lower panel between the pulses of radiation propagating through the plasma (blue curve) and those propagating in vacuum (red dashed curve)  makes it clear that even as the beam is being almost completely attenuated by the plasma,
it propagates at the speed of light and, as far as we can see, it does not broaden.

Extending our calculations with $R_0 \pi T = 0.15$  to
larger $\gamma$ is possible, but the numerics rapidly become more difficult as the radial width of the  pulses narrows like $1/\gamma^3$, rapidly increasing the required dynamical range in momentum space and rapidly making the Fourier transform back to position space more costly.  If it were important to pursue this, however, it could certainly be done.  But, we shall see in Section VI that we have an analytic understanding of the qualitative features in our results, and that based upon this analytic understanding we do not expect any qualitatively new behavior at larger $\gamma$ at the same $R_0$.   
However, it would be very interesting to explore the regime in which
$R_0 \pi T \gg 1$, meaning that $\Omega/\pi T \ll 1$, and  yet $\gamma$ is so large that $\Xi\gg 1$.
In this regime, we expect narrow pulses of radiation that are much more widely separated in the
radial direction than those we have analyzed, say in the right column of Fig.~\ref{blingbling}.
This would allow us to learn more about the response of the strongly coupled plasma 
to a beam of synchrotron radiation since we would be able to watch the plasma for much longer
after one pulse of radiation passed before the next pulse arrived.
Unfortunately, there is at present a serious obstacle to doing such a calculation, more serious than can be overcome simply by increasing the dynamical range of the calculation.  
We saw in Section IV that the helicity 0 gauge invariant quantity $Z$ that we defined in Section IV has gravitational equations of motion that are badly behaved for $q\ll \pi T$, but that because 
the energy density has support only where $q\gtrsim \omega = m\Omega$, with $m$ an integer, this problem only arose for $m=0$ modes --- which we were able to analyze using a different gauge invariant described in Appendix A.  However, if $\Omega\ll\pi T$ then there are many modes 
with $\omega\neq  0$ and $\omega \ll \pi T$ which are important.  We have not found a gauge invariant that yields a tractable numerical analysis for such modes --- the analysis of Appendix A only works for $\omega=0$.   

There are in fact several (related) obstacles to using our calculation to provide definitive
 answers to the question of  where the energy
that is initially in the gluon beam goes as the gluon beam gets attenuated.  The first we have discussed above: we cannot watch the plasma behind one of the pulses of radiation very long before the next pulse comes along and obliterates whatever the previous pulse has left behind.
However, even if a resolution to the technical obstacle that currently precludes addressing this were found, a further obstacle remains.  We are analyzing a scenario in which the quark has been moving in a circle for an infinitely long time meaning that a steady-state in which the energy density  at any position is a periodic function of time has been achieved.  We see in Figs.~\ref{blingbling} and \ref{Beta065blingbling} that the energy density in the beam falls off faster than $1/r^2$ 
at large $r$.  So, the natural first expectation is that the beam heats the plasma up in the range of $r$ over which it gets attenuated --- perhaps it first makes sound waves, but ultimately these too will damp, leaving just a heated region of plasma.  This expectation cannot be correct in a steady-state calculation like the one we have done, since a continual heating up of some region of space blatantly contradicts the steady-state assumption.  So, what actually happens to the energy in our calculation?  We have checked that at sufficiently large $r$ the energy density $\Delta {\cal E}$ is zero.  This means that at sufficiently large $r$, there is an outward flux of energy whose magnitude, averaged over angles, is $P/(4\pi r^2)$ with $P$ the energy lost by the rotating quark per unit time.
This energy flux corresponds to a collective outward flow of the plasma with a velocity, averaged over angles, given by
\begin{equation}
v_{\rm plasma} = \frac{P}{4\pi r^2({\cal E}+p)} 
= \frac{\pi }{2 N_c^2}  \frac{P}{ (\pi T)^2} \frac{1}{(r\,\pi T)^2}
\end{equation}
where we have used the fact that the sum of the energy density and pressure of the plasma in equilibrium is ${\cal E}+p=\pi^2 N_c^2 T^4/2$.  
Since we are working in the large-$N_c$ limit, the velocity $v_{\rm plasma}$ is infinitesimal.  So,  in our steady-state calculation, the energy from the gluon beam ulltimately finds its way into an infinite wavelength mode with infinitesimal amplitude.
A mode like this can equally well be thought of as a sound wave with infinite wavelength and infinitesimal amplitude (i.e. infinitesimal longitudinal velocity) or as a diffusive mode with infinite wavelength.\footnote{In a relativistic plasma, an infinitesimal increase in the temperature in some region must correspond to an infinitesimal increase in the pressure in that region, meaning that it corresponds to sound waves. So, in this case this mode is better thought of as an infinite wavelength sound wave rather than as a diffusive mode.}
This  is the only possible answer to the question of where the energy from the gluon beam ultimately ends up in a steady-state calculation like the one that we have done.  In a sense, this energy flux corresponding to an infinitesimal-velocity 
outward flow of the plasma 
is the closest that a steady-state calculation can come to describing the heating up of a region of the plasma --- which cannot happen in steady-state.

\section{Discussion}

We turn now to a discussion of our results.
Much can be learned about the qualitative features of the results illustrated in Fig.~\ref{blingbling} by studying the quasinormal modes of the AdS-BH spacetime that provides the dual gravitational description of the physics.
In the dual gravitational picture, the moving string excites a full spectrum of gravitational quasinormal modes,
which propagate outwards and eventually get absorbed by the black hole.  The propagation and absorption 
of these quasinormal modes manifests itself on the boundary as the propagation and attenuation of the spirals of energy density shown in Fig.~\ref{blingbling}.
The dispersion relations $\omega(q)$ of the helicity 0 quasinormal modes are given by solutions 
to~\cite{Kovtun:2005ev}
\begin{equation}
\label{qmspectrum}
\det{\mathcal L(\omega,q) } = 0\ ,
\end{equation}
where the linear operator $\mathcal L(\omega,q)$ is defined in Eq.~(\ref{lindef}) and where
the determinant is to be evaluated in the space of functions satisfying the appropriate boundary
conditions that we have described earlier.
Upon introducing a complete basis of functions of $u$ that satisfy the boundary
conditions and then truncating that basis,
%discretizing the $u$ direction using pseudospectral methods, 
$\mathcal L$ becomes 
a matrix and Eq.~(\ref{qmspectrum}) can be solved for $\omega(q)$ numerically.

\begin{figure}[t]
\includegraphics[scale=0.29]{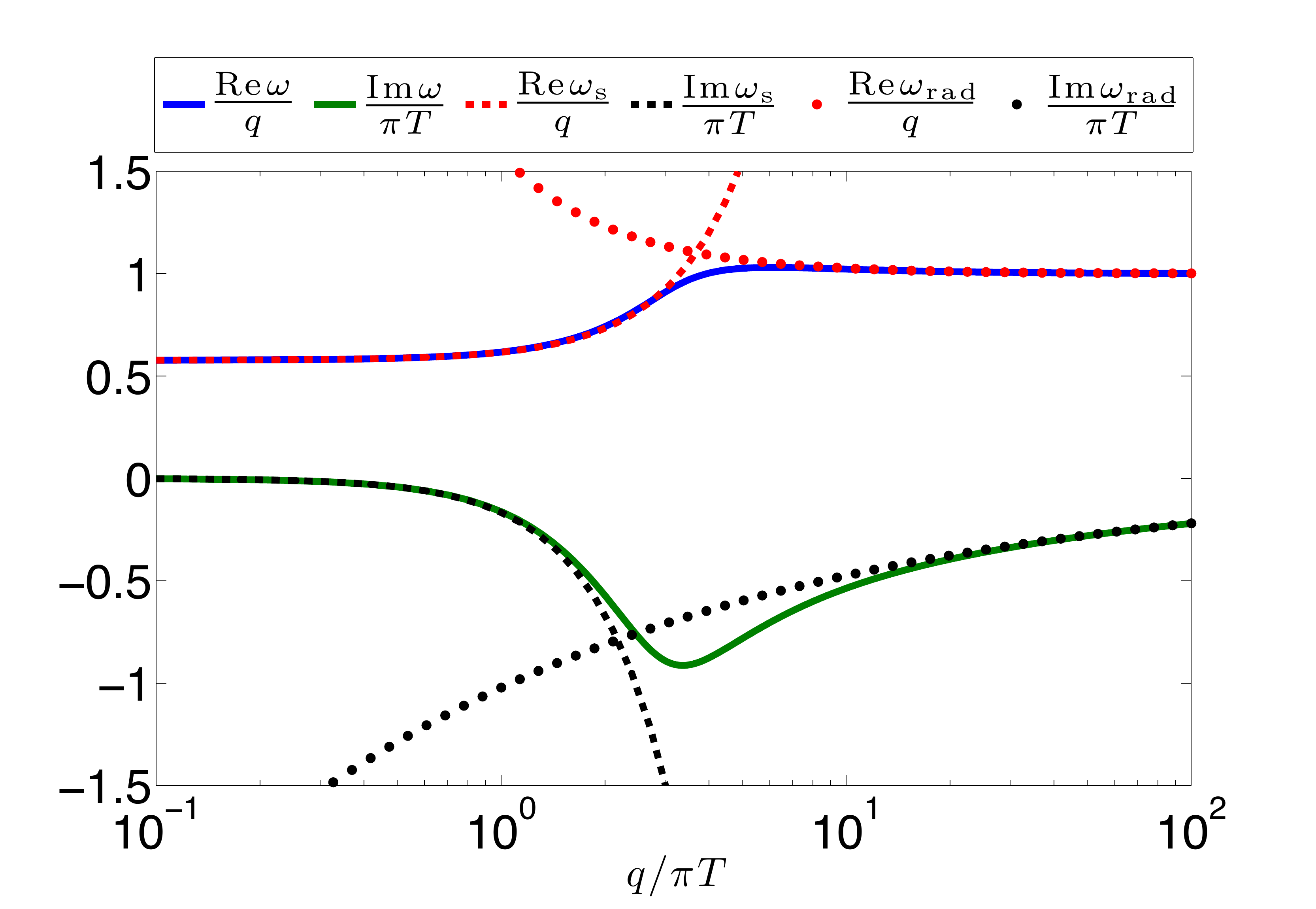}
\caption
    {\label{qmdispersion}
    A plot of the real and imaginary parts of the dispersion relation of the lowest quasinormal mode.  We plot Re$\,\omega / q$ and Im$\,\omega/(\pi T)$ since these ratios are both of order 1. 
    For $q \ll \pi T$ the 
    dispersion relation is that of sound waves whose dispersion relation is given up to order $q^3$ 
by Eq.~({\protect\ref{SoundDispersion}}), plotted as dashed lines in the figure.    
        For $q \gg \pi T$ the dispersion relation is that of waves which 
    propagate at the speed of light.  The large-$q$ asymptotic expression ({\protect\ref{lightlike}}) that we have obtained by fitting the results in this figure is plotted as the dotted lines.
     }
\end{figure}

Fig.~\ref{qmdispersion} shows a plot of the quasinormal mode dispersion relation for the 
lowest quasinormal mode (\textit{i.e.} the one with the smallest imaginary part).  For $q \ll \pi T$ the dispersion relation
has the asymptotic form expected for the hydrodynamics of any conformal fluid~\cite{Baier:2007ix}
\begin{equation}
\omega_s(q) = c_s q - i \Gamma q^2  + \frac{\Gamma}{c_s}\left( c_s^2\, \tau_\Pi  - \frac{\Gamma}{2}\right) q^3 + {\cal O}(q^4)\ ,
\label{SoundDispersion}
\end{equation}
where in ${\cal N}=4$ SYM theory, with its classical gravity dual, all the constants are known analytically: the low-$q$ speed of sound is $c_s=1/\sqrt{3}$, the sound attenuation constant $\Gamma$ is given by
$\pi T \,\Gamma = 1/6$, and the relaxation time $\tau_\Pi$ is 
given by $\pi T \tau_\Pi = (2 - \log 2)/2$.  
These modes represent 
propagating sound waves which attenuate over a timescale 
\begin{equation}
t_s^{\rm damping} \sim \frac{1}{\Gamma q^2}\ .
\label{SoundDamping}
\end{equation}
The dispersion relation (\ref{SoundDispersion}) is plotted in Fig.~\ref{qmdispersion}; it describes the full dispersion relation very well for $q\lesssim 2 \pi T$.  This supports our observation that the waves in the left column of Fig.~\ref{blingbling} are sound waves.
Since these waves are not monochromatic (and since in the dual gravitational description
they are not described solely by the lowest quasinormal mode) 
they cannot be compared quantitatively 
to (\ref{SoundDispersion}), but their velocity is as (\ref{SoundDispersion}) predicts for $q\sim 1.2\, \pi T$, which is comparable to the $q\sim 1.35 \,\pi T$ obtained from their peak-to-peak 
wavelength.    Using $q\sim 1.2\, \pi T$ in (\ref{SoundDamping}) predicts a sound attenuation timescale $(\Delta t)_{\rm sound} \sim 4.5/(\pi T) \sim 30\, R_0$, 
which is comparable to but a little shorter than 
the exponential decay time for the amplitude of the waves in the left column of Fig.~\ref{blingbling}, which is closer to $40 \,R_0$.  So, although a quantitative comparison is not possible, it does seem that the low-$q$ regime of the dispersion relation in Fig.~\ref{qmdispersion}  that describes sound waves does a reasonable job of capturing the qualitative features of the waves seen in the left column of Fig.~\ref{blingbling}.

The dispersion relations of the higher quasinormal modes (those with more negative imaginary parts) can also be determined by solving Eq.~(\ref{qmspectrum}) numerically.
At $q \ll \pi T$ they approach the asymptotic form $\omega = (\tilde a - i \tilde b) \pi T$ where $\tilde a$ and $\tilde b$ are mode-dependent ${\cal O}(1)$ constants, with values that are larger and larger for higher and higher modes.  (For the lowest quasinormal mode, $\tilde a=\tilde b=0$.)
At low $q$, disturbances of the plasma described by  higher quasinormal modes attenuate on a timescale of order $1/(\tilde b \,\pi T)$ that is much shorter than that for
the sound waves described by the lowest quasinormal mode, namely (\ref{SoundDamping}).

Let us turn now to  $q \gg \pi T$. By fitting to our results for the dispersion relation for the lowest quasinormal mode, we find that
in this regime the dispersion relation takes the asymptotic form 
\begin{equation}
\label{lightlike}
\omega_{\rm rad} = q + \pi T (a-i b) \left (\frac{ \pi T}{q} \right )^{1/3} + \ldots,
\end{equation}
as argued for on general grounds in Ref.~\cite{Festuccia:2008zx},
where we find $a \approx 0.58$ and  $b \approx 1.022 $.  
At $q \gg \pi T$ the dispersion relations of all quasinormal modes approach the asymptotic form
(\ref{lightlike}), with $a$ and $b$ mode-dependent ${\cal O}(1)$ constants, again with values that are larger and larger for higher and higher modes.  Therefore, generically the high $q$ modes propagate at close to the speed of light
and attenuate over a time-scale 
\begin{equation}
t_{\rm rad}^{\rm damping} \sim \frac{1}{\pi T b} \left(\frac{q}{\pi T}\right)^{1/3}\ .
\label{LightDamping}
\end{equation}
where we shall use the value $b\approx 1.022$ from the lowest quasinormal mode in making estimates, keeping in mind that if the contribution of higher quasinormal modes were important this would increase the effective $b$ somewhat.  The fact that the pulses of energy in 
Fig.~\ref{blingbling} are far from being monochromatic waves introduces a larger uncertainty into our discussion than does not knowing how much the higher quasinormal modes contribute.

We have plotted the large-$q$ asymptotic expression (\ref{lightlike}) for the dispersion relation
for the lowest quasinormal mode 
in Fig.~\ref{qmdispersion}, and we see that it describes the full result very well for $q\gtrsim 20\,\pi T$, and has the right shape at a qualitative level down to about $q\sim 5\,\pi T$.  
This is consistent with our observation that the narrow pulses of synchrotron radiation in the middle column, where the pulses have a full width at half maximum (FWHM)
$\sim  2.5 \,R_0$ corresponding very roughly to $q\sim 6\,  \pi T$, and the right column, where the pulses have a FWHM $\sim R_0$ corresponding very roughly to $q\sim 15\, \pi T$,  propagate outwards at the speed of light.    Converting the widths of these pulses into estimates of $q$ is very rough because the pulses are neither sinusoidal nor Gaussian.  If we nevertheless try substituting $q\sim 15\, \pi T$ into (\ref{LightDamping}) we find that it predicts 
$t_{\rm rad}^{\rm damping}\sim 16\,R_0$, which is roughly half the exponential decay time for the 
amplitude of the waves in the lower-right panel of Fig.~\ref{blingbling}.  Again, quantitative comparison is not possible, but inferences drawn from the large-$q$ dispersion relation for the lowest quasinormal mode (\ref{lightlike}) is at least in the right ballpark.  

We can also compare $t_{\rm rad}^{\rm damping}$ for the radiation emitted by the quark moving with $\beta=0.5$ in the right column of Fig.~\ref{blingbling} with that emitted by the quark moving with $\beta=0.65$ in Fig.~\ref{Beta065blingbling}. Since the pulses of synchtron radiation narrow proportional to $1/\gamma^3$ as $\beta$ increases, Eq.~(\ref{LightDamping}) predicts that 
$t_{\rm rad}^{\rm damping}$ should increase proportional to $\gamma$, namely by about 14\% in going from $\beta=0.5$ to $\beta=0.65$.  It is hard to define and extract a damping time from the figures with sufficient accuracy to test this prediction quantitatively but, again, it is in the right ballpark.

The qualitative prediction from (\ref{LightDamping}) that narrower pulses, with higher $q$, can penetrate farther into the strongly coupled quark-gluon plasma
is apparent in comparing the $\theta$-dependence of the results in the top-right panel of Fig.~\ref{blingbling} with the $T=0$ results in Fig.~\ref{ZeroTRad}.  We see that in vacuum the pulses of synchrotron radiation get broader and broader as you decrease $\theta$ from $\pi/2$, going from the equator toward the poles.  And, indeed, just as (\ref{LightDamping}) predicts we see in Fig.~\ref{blingbling} that in the quark-gluon plasma the radiation heading in polar directions is quenched much more quickly than that radiated in narrower pulses near the equator.

We can also use the quasinormal mode dispersion relation to understand why the pulses do not broaden significantly in the radial direction as they propagate.  The increase in the width of a pulse as it propagates for a time $t$ is  $\sim t\,  \Delta q \, d^2 \omega/dq^2$, where $\Delta q$ is the width of the pulse in $q$-space.  Taking $\Delta q \sim q$ and using the large-$q$ dispersion relation (\ref{lightlike}), we find that after the radiation damping time given by (\ref{LightDamping}) the pulse should have broadened by $\sim 4a/(9b q)$.  If the pulse had a Gaussian profile, this would correspond to broadening by about 10\% of the original FWHM 
of the pulse.  
So, the quasinormal mode dispersion relation predicts that by the time the pulses have been significantly attenuated, they should have broadened by an amount that is parametrically of order their initial width, but smaller by a significant numerical factor.  It is therefore not surprising that we see no significant broadening in Fig.~\ref{blingbling}.

%We now use the quasinormal mode dispersion relation (\ref{lightlike}) to study the attenuation and broadening of the 
%light-like spirals shown in Fig.~\ref{blingbling}.  For $\Xi \gg1$, locally the radiation 
%emitted by the quark resembles synchrotron radiation at $T = 0$.  At $T = 0$ the spirals have a 
%radial thickness $\Delta \sim R_0/\gamma^3$ and are localized about $\theta = \pi/2$ with a characteristic width $\delta \theta \sim 1/\gamma$ [[our old paper]].
%It therefore follows that the characteristic momentum which contributes to the spiral is $q \sim \gamma^3/R_0$.   From the dispersion relation
%({\ref{lightlike}) the corresponding quasinormal modes attenuate 
%over a time $\Delta v_{\rm att} \sim \frac{\gamma}{\pi T b} \frac{1}{(\pi T R_0)^{1/3}}$.  
%Moreover, from the dispersion relation (\ref{lightlike}) we see that 
%large $q$ modes propagate at speed $1 - a \left(  \pi T/q \right )^{4/3}$.  It therefore follows that over the time $\Delta v_{\rm att}$ the width of the spirals
%will have increased by an amount $\sim a \left( \frac{ \pi T}{q} \right )^{4/3} \Delta v_{\rm att} \sim \frac{a}{b} \frac{R_0}{\gamma^3}$.
%Thus the increase of the width over the attenuation time scale is parametrically the same size as the original width!  We therefore conclude that in the $\gamma \to \infty$ limit
%the light-like spirals penetrate a distance $\Delta r \sim \frac{\gamma}{\pi T b} \frac{1}{(\pi T R_0)^{1/3}}$ before attenuating and always have a radial thickness $\Delta \sim R_0 /\gamma^3$.

By this point we have understood many of the most interesting features of our results qualitatively, and even semi-quantitatively, by analyzing the quasinormal mode dispersion relations.
This gives us confidence that no new qualitative phenomena emerge for narrower pulses (higher $q$; e.g. from a rotating quark with larger $\gamma$) than we have been able to study, since it is clear that our results at $\beta=0.5$ and $\beta=0.65$ are already exploring the high-$q$ regime of the dispersion relation in Fig.~\ref{qmdispersion}, where the asymptotic 
expression (\ref{lightlike}) is a good guide.
It is also important to stress that the quasinormal mode frequencies 
are determined entirely by ${\cal L}$, from the left-hand side of (\ref{Zeq}).
This means they reflect properties of the strongly coupled plasma itself and have nothing
to do with the source on the right-hand side of (\ref{Zeq}), namely the rotating quark.
Given that we have been able to use the quasinormal mode dispersion relations
so successfully to understand the propagation, rate of attenuation and lack of broadening of a beam of gluons, 
we are confident that these phenomena are are independent of how the beam of gluons
is created.

Finally, we can use our understanding of the quasinormal mode dispersion relations to clarify
the 
distinction between the middle column of Fig.~\ref{blingbling} --- in which we see the pulse of
radiation shedding a sound wave --- and the cases where we don't see this (like the right column of Fig.~\ref{blingbling} and Fig.~\ref{Beta065blingbling}.)  As we described at the end of Section V, the fact that we have done a steady-state calculation makes it difficult to use our results directly to draw conclusions about what our pulse of radiation leaves behind.  However, we can use our understanding of the quasinormal modes to answer this question at least qualitatively.  
Suppose that we could move the quark on some trajectory such that it radiates one pulse of synchrotron radiation (i.e. one turn of the spiral) and then no more;  what would happen to this single pulse of strongly coupled radiation as it propagates outward through the strongly coupled plasma?  The dual gravitational description of this radiation would be governed
by (\ref{Zeq}), with the same ${\cal L}$ and hence the same quasinormal modes as in our analysis but with a different source $S$, localized along the world sheet of the string hanging down
from the quark that excited the single pulse of radiation.  As long as we look only at distances greater than of order $1/(\pi T)$ away from the location of the string, the disturbance of the plasma must be described by a pulse of short wavelength radiation with the dispersion relation (\ref{lightlike}) that moves at the speed of light and is attenuated on timescales (\ref{LightDamping})
as well as long wavelength sound waves with the dispersion relation (\ref{SoundDispersion}) that propagate outward at the speed of sound and are attenuated on timescales (\ref{SoundDamping}).  Since these sound waves move more slowly, the pulse of radiation leaves them behind --- shedding them as we see in the middle column of Fig.~\ref{blingbling}.   
The same would happen for shorter wavelength pulses as in the right column 
of Fig.~\ref{blingbling} or as in Fig.~\ref{Beta065blingbling}, but in these cases in our steady-state calculation the next pulse of synchrotron radiation arrives before we can see the sound waves being left behind.  
In the hypothetical case of a single short wavelength pulse, the short wavelength pulse itself will get far ahead of the sound waves it has left behind as it is attenuated only on the long timescale (\ref{LightDamping}).  By the time the short wavelength pulse has damped away, the 
sound waves that it shed will be far behind and will have become sound waves with very small $q$, meaning long wavelengths and small amplitudes, since (\ref{SoundDamping}) tells us that these are the sound modes that last the longest. 
These are
represented in the steady-state calculation by the
outward-going energy flux at infinite wavelength and infinitesimal amplitude 
that we described at the end of Section V.  We now see that the distinction between the middle column of Fig. ~\ref{blingbling} on the one hand and the right column of Fig.~\ref{blingbling} and Fig.~\ref{Beta065blingbling} on the other hand is that in the former case the pulse of radiation is never well-separated from the sound waves that it leaves behind --- if a significant
pulse of radiation remains, the sound waves are not far behind it and have themselves not yet been attenuated --- while in the latter case the pulse of radiation gets far ahead of the sound waves and the sound waves ``thermalize'' (which means increase in wavelength and decrease in amplitude) before the pulse of radiation has been attenuated.

\section{From Quenching a Beam of Strongly Coupled Gluons to Jet Quenching}

There are many qualitative similarities between the quenching of the 
beam of strongly coupled synchrotron radiation in the strongly coupled ${\cal N}=4$ SYM plasma
that we have studied and the quenching of jets in heavy ion collisions at the LHC and RHIC:
\begin{itemize}
\item
As our beam of gluons propagates through the plasma, losing a significant fraction of its energy, it does not spread in 
angle.  Jets in heavy ion collisions at the LHC lose a significant fraction of their energy but, within current experimental errors, do not broaden in angle and do not get deflected in their direction.
\item
As our beam of gluons propagates through the plasma, and is significantly attenuated, it
does not spread in the direction along which it propagates.  In other words, the momenta of the gluons making up the beam do not decrease even as the beam loses a significant fraction of its energy.  Similarly, jets in heavy ion collisions at the LHC have fragmentation functions (i.e. distributions of the momenta of the particles making up the jet) that are unmodified by
propagation through the strongly coupled plasma produced in the collisions except 
by virtue of the overall reduction in the energy of the jet.  
\item
In the case of a  beam made up of gluons whose wavelength is not too short, as in the middle column of Fig.~\ref{blingbling}, it is possible for a significantly attenuated pulse of radiation to be followed close on its heels by a significant sound wave, trailing behind it.  In contrast, once a beam made 
up of shorter wavelength gluons, as in the right
column of Fig.~\ref{blingbling} or in Fig.~\ref{Beta065blingbling}, 
is significantly attenuated, the sound waves that it left behind are far behind it and have 
themselves thermalized. In our steady-state calculation,  they have ended up in the infinite wavelength,
infinitesimal velocity, mode that we described at the end of Section V.
The behavior of these shorter wavelength pulses, losing their energy
to a mode in which the collective velocity of the fluid is infinitesimal,
suggests the observation that jets at 
the LHC lose their energy to soft particles at all angles relative to the jet direction.  
The behavior of the longer wavelength pulses suggests that
there may be a regime of jet energies, analogous to the beam of radiation in the middle column in Fig.~\ref{blingbling}, in which at a time when the jet itself has been attenuated significantly
it is followed by a significant pulse of sound waves moving in the same direction as the jet.
If the jets accessible in RHIC collisions, with energies in the 20-40 GeV range, are in this regime, this would indicate that they should lose their energy to soft particles that are correlated in angle with the jet direction, consistent with preliminary indications from RHIC data.
\end{itemize}
Comparisons along these lines will never be more than qualitative, since the beam of strongly coupled radiation that we have analyzed is not a jet.  However, these multiple qualitative resonances
between jet quenching in heavy ion collisions and the quenching of the beam of strongly
coupled radiation that we have analyzed support the prospect that jet quenching is a strongly coupled phenomenon, even for the few-hundred-GeV jets produced at the LHC.  
If this is so, what is to be done next? Further directions include:
\begin{itemize}
\item
Finding a helicity 0 gauge invariant that makes the regime with $R_0\pi T \gg 1$ and $\gamma$ large enough that $\Xi\gg 1$ accessible, since this would give much more time for the region of plasma through which a pulse of radiation has passed to be analyzed before the next pulse hits
and thus would permit a more definitive analysis of where the energy lost by the quenched beam 
ends up.
\item
Finding ways to make a beam of strongly coupled radiation other than via synchrotron radiation, and in particular finding a way to make such a beam pointing in a pair of  fixed directions rather than rotating.  For example, could multiple quarks moving in concert be engineered so as to behave like a phased array of antennas, generating back-to-back beams of strongly coupled radiation?
\item
 Reanalyzing the stopping of a light quark or gluon, as in 
 Refs.~\cite{Gubser:2008as,Chesler:2008wd,Chesler:2008uy}, should give another means of
 accessing many of the questions we have addressed.   And, these analyses find a stopping distance for energetic light quarks that is proportional to $E^{1/3}$, with $E$ the energy of the quark.
 Our result for the distance scale over which our beam of gluons is quenched, namely $t_{\rm rad}^{\rm damping}  \sim q^{1/3}/(\pi T)^{4/3}$ where $q$ is the typical wave vector of the gluons
 in the beam, has the same parametric dependence as the light-quark stopping distance, adding considerable robustness to both results.
 There have been some first steps taken  to compare 
 this relationship between quark or gluon energy or wave vector and stopping or quenching distance with heavy ion collision 
 data~\cite{Horowitz:2011cv,Bathe:2011xx}.  
 Our results further motivate these efforts.

 \item
It is also interesting to note that in the calculation of Ref.~\cite{Chesler:2008uy} a very high energy quark that loses, say, half of its energy leaves most of that energy far behind it while a lower energy quark that loses the same amount of energy and comes almost to rest is never well-separated from the energy it has lost.  Although described in quite different terms, this is reminiscent of the distinction between the middle and right panels of Fig.~\ref{blingbling}, and perhaps of the possible 
distinction between jet quenching at the LHC and at RHIC.   This distinction may also be
seen in the analysis of the energy loss of a heavy quark being dragged through
the strongly coupled plasma~\cite{Gubser:2006bz,Herzog:2006gh}, in which the world sheet of the string trailing behind the dragging quark features a world-sheet
 horizon~\cite{CasalderreySolana:2006rq,Gubser:2006nz,CasalderreySolana:2007qw}, located near the spacetime horizon for a low velocity quark and located closer and closer to the boundary for quarks moving more and more relativistically.  
One possible interpretation of the world-sheet horizon is that the portion of the string between it and the boundary describes (ultraviolet) modes in the gauge theory that propagate along with the heavy quark while the portion of the string between the world-sheet horizon and the spacetime horizon describe the disturbance of the plasma that the quark leaves 
behind~\cite{Chernicoff:2008sa,Dominguez:2008vd,Xiao:2008nr,Beuf:2008ep}.  This interpretation predicts that a higher energy heavy quark will be accompanied only by shorter wavelength modes of the gauge theory while a
lower energy heavy quark will be accompanied by softer gauge theory modes.  
Although described in terms that are different yet again, this is again reminiscent of the
possibility that only lower energy quenched jets will be accompanied by soft particles correlated in
angle with the jet direction.
All these on-the-surface quite distinct approaches to 
jet quenching point in the same direction, suggesting that the distinction manifest in Fig.~\ref{blingbling}
between an attenuated high energy jet that has
left its sound waves far behind and an attenuated lower energy jet that has a pulse of sound close on its heels may be generic to jet quenching in strongly coupled plasma.  The connections between these different approaches must be explored and developed.

 \item
 Our results also motivate further effort to determine whether the soft particles 
 in a heavy ion collision corresponding to the energy lost by a quenched jet 
 really are more correlated with the jet direction at lower jet energies.  If so,
 this will point to lower jet energies as the place to look for the hydrodynamic
 response of the plasma.
\item
On a more theoretical note, it would be very interesting to repeat our analyses in a
nonconformal strongly coupled gauge theory plasma with a dual gravitational description. 
  \end{itemize}

\acknowledgments

We acknowledge helpful conversations with Christiana Athanasiou, Andreas Karch, Hong Liu, Berndt Muller, Dominik Nickel and Urs Wiedemann. 
The work of PC is supported by a Pappalardo Fellowship in Physics at MIT. 
This research was
supported in part by the DOE Office of Nuclear Physics under grant \#DE-FG02-94ER40818.

\appendix

\section{Zero Mode Gauge Invariant}

The helicity 0 gauge invariant $Z$ that we defined in Section IV is inconvenient at 
$m=\omega=0$ because the equations for it have apparent divergences that preclude their numerical solution.  In the static case with $\omega=0$
a different helicity 0 gauge invariant
\begin{align}
Z_0 \equiv &q^2 \H_{00} + \frac{q^2}{2} \left(2 - f \right )(\H_{ii} - \H_{qq})
\label{Z0Defn}
\end{align}
proves convenient instead.\footnote{The definition of $Z_0$ that we have given is valid only at $\omega=0$ but this definition can be extended to nonzero $\omega$. Upon doing so, the expression (\ref{Z0Defn}) becomes much lengthier and, furthermore,
one discovers that it features ratios of terms that both vanish at certain values of $u$ between 0 and 1.  Although $Z_0$ is in principle well-behaved at these values of $u$, this feature makes $Z_0$ impractical for numerical calculations when $\omega\neq 0$. No such difficulties arise at $\omega=0$, which is the only case for which we use $Z_0$.}
As one can easily verify, $Z_0$ is invariant under time-independent infinitesimal diffeomorphisms
and satisfies the equation of motion
\begin{equation}
\mathcal L_0 Z_0 = \kappa_5^2 S_0,
\end{equation}
where the linear operator $\mathcal L_0$ is given by
\begin{align}
\mathcal L_0 = f \frac{d^2}{du^2} - \frac{9 - 8u^4 + 7 u^8}{u (3 - u^4)} \frac{d}{du} 
- q^2 + \frac{16 u^6}{3 - u^4}
\end{align}
and the source $S_0$ is given by
\begin{align}
\nonumber
S_0 =& -2 q^2 \t_{00}-\frac{q^2(3 {-} u^4)}{3}\t_{ii} + q^2(1{+}u^4) \t_{qq} - \frac{32u^3 i q }{3 {-} u^4} \t_{0q}\\
&- \frac{8}{3} q^2 u^4 \t_{05} + \frac{4}{3} q^2 u^4 f \t_{55} + \frac{32 u^3 i q f}{3 - u^4} \t_{q5}.
\end{align}
Near the boundary $Z_0$ has the following expansion
\begin{equation}
Z_0 = Z^0_{(3)} u^3 + Z^0_{(4)} u^4 + \dots.
\end{equation}
The zero mode of the boundary energy density is related to $Z^0_{(4)} $ via~\cite{Chesler:2007sv}
\begin{align}
\mathcal E = \frac{4}{3 q^2 \kappa^2_5} Z^0_{(4)} + \mathcal A,
\end{align}
where 
\begin{align}
\mathcal A \equiv -\frac{\sqrt{\lambda}}{2 \pi} \,\frac{R_3 \sqrt{1 {-} \Omega^2 R_0^2}}{q R_0 } 
\,\frac{\partial j_{\ell}(q R_0)}{\partial q}\, Y_{\ell 0}({\textstyle \frac{\pi}{2}},0),
\end{align}
with $R_3 \equiv \frac{1}{6} \lim_{u \to 0} \partial_u^3 R$ and $R_0 = \lim_{u \to 0} R$.

\vfill\eject

\bibliography{refs}

\begin{thebibliography}{69}
\expandafter\ifx\csname natexlab\endcsname\relax\def\natexlab#1{#1}\fi
\expandafter\ifx\csname bibnamefont\endcsname\relax
  \def\bibnamefont#1{#1}\fi
\expandafter\ifx\csname bibfnamefont\endcsname\relax
  \def\bibfnamefont#1{#1}\fi
\expandafter\ifx\csname citenamefont\endcsname\relax
  \def\citenamefont#1{#1}\fi
\expandafter\ifx\csname url\endcsname\relax
  \def\url#1{\texttt{#1}}\fi
\expandafter\ifx\csname urlprefix\endcsname\relax\def\urlprefix{URL }\fi
\providecommand{\bibinfo}[2]{#2}
\providecommand{\eprint}[2][]{\url{#2}}

\bibitem[{\citenamefont{Aad et~al.}(2010)}]{Aad:2010bu}
\bibinfo{author}{\bibfnamefont{G.}~\bibnamefont{Aad}} \bibnamefont{et~al.}
  (\bibinfo{collaboration}{ATLAS Collaboration}), \bibinfo{journal}{Phys. Rev.
  Lett.} \textbf{\bibinfo{volume}{105}}, \bibinfo{pages}{252303}
  (\bibinfo{year}{2010}), \eprint{1011.6182}.

\bibitem[{\citenamefont{Chatrchyan et~al.}(2011)}]{Chatrchyan:2011sx}
\bibinfo{author}{\bibfnamefont{S.}~\bibnamefont{Chatrchyan}}
  \bibnamefont{et~al.} (\bibinfo{collaboration}{CMS Collaboration}),
  \bibinfo{journal}{Phys. Rev.} \textbf{\bibinfo{volume}{C84}},
  \bibinfo{pages}{024906} (\bibinfo{year}{2011}), \eprint{1102.1957}.

\bibitem[{\citenamefont{Angerami}(2011)}]{Angerami:2011is}
\bibinfo{author}{\bibfnamefont{A.}~\bibnamefont{Angerami}}
  (\bibinfo{collaboration}{ATLAS Collaboration}) (\bibinfo{year}{2011}),
  \eprint{1108.5191}.

\bibitem[{\citenamefont{Wyslouch}(2011)}]{Collaboration:2011yma}
\bibinfo{author}{\bibfnamefont{B.}~\bibnamefont{Wyslouch}}
  (\bibinfo{collaboration}{CMS Collaboration}) (\bibinfo{year}{2011}),
  \eprint{1107.2895}.

\bibitem[{\citenamefont{Roland}(2011)}]{Collaboration:2011cs}
\bibinfo{author}{\bibfnamefont{C.}~\bibnamefont{Roland}}
  (\bibinfo{collaboration}{CMS Collaboration}) (\bibinfo{year}{2011}),
  \eprint{1107.3106}.

\bibitem[{\citenamefont{Steinberg}(2011)}]{Steinberg:2011dj}
\bibinfo{author}{\bibfnamefont{P.}~\bibnamefont{Steinberg}}
  (\bibinfo{collaboration}{ATLAS Collaboration}) (\bibinfo{year}{2011}),
  \eprint{1107.2182}.

\bibitem[{\citenamefont{Gyulassy and Wang}(1994)}]{Gyulassy:1993hr}
\bibinfo{author}{\bibfnamefont{M.}~\bibnamefont{Gyulassy}} \bibnamefont{and}
  \bibinfo{author}{\bibfnamefont{X.-n.} \bibnamefont{Wang}},
  \bibinfo{journal}{Nucl. Phys.} \textbf{\bibinfo{volume}{B420}},
  \bibinfo{pages}{583} (\bibinfo{year}{1994}), \eprint{nucl-th/9306003}.

\bibitem[{\citenamefont{Baier et~al.}(1997)\citenamefont{Baier, Dokshitzer,
  Mueller, Peigne, and Schiff}}]{Baier:1996sk}
\bibinfo{author}{\bibfnamefont{R.}~\bibnamefont{Baier}},
  \bibinfo{author}{\bibfnamefont{Y.~L.} \bibnamefont{Dokshitzer}},
  \bibinfo{author}{\bibfnamefont{A.~H.} \bibnamefont{Mueller}},
  \bibinfo{author}{\bibfnamefont{S.}~\bibnamefont{Peigne}}, \bibnamefont{and}
  \bibinfo{author}{\bibfnamefont{D.}~\bibnamefont{Schiff}},
  \bibinfo{journal}{Nucl. Phys.} \textbf{\bibinfo{volume}{B484}},
  \bibinfo{pages}{265} (\bibinfo{year}{1997}), \eprint{hep-ph/9608322}.

\bibitem[{\citenamefont{Zakharov}(1997)}]{Zakharov:1997uu}
\bibinfo{author}{\bibfnamefont{B.}~\bibnamefont{Zakharov}},
  \bibinfo{journal}{JETP Lett.} \textbf{\bibinfo{volume}{65}},
  \bibinfo{pages}{615} (\bibinfo{year}{1997}), \eprint{hep-ph/9704255}.

\bibitem[{\citenamefont{Wiedemann}(2000)}]{Wiedemann:2000za}
\bibinfo{author}{\bibfnamefont{U.~A.} \bibnamefont{Wiedemann}},
  \bibinfo{journal}{Nucl. Phys.} \textbf{\bibinfo{volume}{B588}},
  \bibinfo{pages}{303} (\bibinfo{year}{2000}), \eprint{hep-ph/0005129}.

\bibitem[{\citenamefont{Guo and Wang}(2000)}]{Guo:2000nz}
\bibinfo{author}{\bibfnamefont{X.-F.} \bibnamefont{Guo}} \bibnamefont{and}
  \bibinfo{author}{\bibfnamefont{X.-N.} \bibnamefont{Wang}},
  \bibinfo{journal}{Phys. Rev. Lett.} \textbf{\bibinfo{volume}{85}},
  \bibinfo{pages}{3591} (\bibinfo{year}{2000}), \eprint{hep-ph/0005044}.

\bibitem[{\citenamefont{Gyulassy et~al.}(2001)\citenamefont{Gyulassy, Levai,
  and Vitev}}]{Gyulassy:2000er}
\bibinfo{author}{\bibfnamefont{M.}~\bibnamefont{Gyulassy}},
  \bibinfo{author}{\bibfnamefont{P.}~\bibnamefont{Levai}}, \bibnamefont{and}
  \bibinfo{author}{\bibfnamefont{I.}~\bibnamefont{Vitev}},
  \bibinfo{journal}{Nucl. Phys.} \textbf{\bibinfo{volume}{B594}},
  \bibinfo{pages}{371} (\bibinfo{year}{2001}), \eprint{nucl-th/0006010}.

\bibitem[{\citenamefont{Casalderrey-Solana
  et~al.}(2011{\natexlab{a}})\citenamefont{Casalderrey-Solana, Milhano, and
  Wiedemann}}]{CasalderreySolana:2010eh}
\bibinfo{author}{\bibfnamefont{J.}~\bibnamefont{Casalderrey-Solana}},
  \bibinfo{author}{\bibfnamefont{J.~G.} \bibnamefont{Milhano}},
  \bibnamefont{and} \bibinfo{author}{\bibfnamefont{U.~A.}
  \bibnamefont{Wiedemann}}, \bibinfo{journal}{J. Phys. G}
  \textbf{\bibinfo{volume}{G38}}, \bibinfo{pages}{035006}
  (\bibinfo{year}{2011}{\natexlab{a}}), \eprint{1012.0745}.

\bibitem[{\citenamefont{Qin and Muller}(2011)}]{Qin:2010mn}
\bibinfo{author}{\bibfnamefont{G.-Y.} \bibnamefont{Qin}} \bibnamefont{and}
  \bibinfo{author}{\bibfnamefont{B.}~\bibnamefont{Muller}},
  \bibinfo{journal}{Phys. Rev. Lett.} \textbf{\bibinfo{volume}{106}},
  \bibinfo{pages}{162302} (\bibinfo{year}{2011}), \eprint{1012.5280}.

\bibitem[{\citenamefont{Young et~al.}(2011{\natexlab{a}})\citenamefont{Young,
  Schenke, Jeon, and Gale}}]{Young:2011qx}
\bibinfo{author}{\bibfnamefont{C.}~\bibnamefont{Young}},
  \bibinfo{author}{\bibfnamefont{B.}~\bibnamefont{Schenke}},
  \bibinfo{author}{\bibfnamefont{S.}~\bibnamefont{Jeon}}, \bibnamefont{and}
  \bibinfo{author}{\bibfnamefont{C.}~\bibnamefont{Gale}}
  (\bibinfo{year}{2011}{\natexlab{a}}), \eprint{1103.5769}.

\bibitem[{\citenamefont{Zapp et~al.}(2011)\citenamefont{Zapp, Stachel, and
  Wiedemann}}]{Zapp:2011ya}
\bibinfo{author}{\bibfnamefont{K.~C.} \bibnamefont{Zapp}},
  \bibinfo{author}{\bibfnamefont{J.}~\bibnamefont{Stachel}}, \bibnamefont{and}
  \bibinfo{author}{\bibfnamefont{U.~A.} \bibnamefont{Wiedemann}},
  \bibinfo{journal}{JHEP} \textbf{\bibinfo{volume}{1107}}, \bibinfo{pages}{118}
  (\bibinfo{year}{2011}), \eprint{1103.6252}.

\bibitem[{\citenamefont{Casalderrey-Solana and
  Iancu}(2011)}]{CasalderreySolana:2011rz}
\bibinfo{author}{\bibfnamefont{J.}~\bibnamefont{Casalderrey-Solana}}
  \bibnamefont{and} \bibinfo{author}{\bibfnamefont{E.}~\bibnamefont{Iancu}},
  \bibinfo{journal}{JHEP} \textbf{\bibinfo{volume}{1108}}, \bibinfo{pages}{015}
  (\bibinfo{year}{2011}), \eprint{1105.1760}.

\bibitem[{\citenamefont{Qin}(2011)}]{Qin:2011ng}
\bibinfo{author}{\bibfnamefont{G.-Y.} \bibnamefont{Qin}}
  (\bibinfo{year}{2011}), \eprint{1107.0631}.

\bibitem[{\citenamefont{Casalderrey-Solana
  et~al.}(2011{\natexlab{b}})\citenamefont{Casalderrey-Solana, Milhano, and
  Wiedemann}}]{CasalderreySolana:2011rq}
\bibinfo{author}{\bibfnamefont{J.}~\bibnamefont{Casalderrey-Solana}},
  \bibinfo{author}{\bibfnamefont{J.}~\bibnamefont{Milhano}}, \bibnamefont{and}
  \bibinfo{author}{\bibfnamefont{U.}~\bibnamefont{Wiedemann}}
  (\bibinfo{year}{2011}{\natexlab{b}}), \eprint{1107.1964}.

\bibitem[{\citenamefont{Coleman-Smith et~al.}(2011)\citenamefont{Coleman-Smith,
  Qin, Bass, and Muller}}]{ColemanSmith:2011rw}
\bibinfo{author}{\bibfnamefont{C.}~\bibnamefont{Coleman-Smith}},
  \bibinfo{author}{\bibfnamefont{G.-Y.} \bibnamefont{Qin}},
  \bibinfo{author}{\bibfnamefont{S.}~\bibnamefont{Bass}}, \bibnamefont{and}
  \bibinfo{author}{\bibfnamefont{B.}~\bibnamefont{Muller}}
  (\bibinfo{year}{2011}), \eprint{1108.5662}.

\bibitem[{\citenamefont{Young et~al.}(2011{\natexlab{b}})\citenamefont{Young,
  Jeon, Gale, and Schenke}}]{Young:2011va}
\bibinfo{author}{\bibfnamefont{C.}~\bibnamefont{Young}},
  \bibinfo{author}{\bibfnamefont{S.}~\bibnamefont{Jeon}},
  \bibinfo{author}{\bibfnamefont{C.}~\bibnamefont{Gale}}, \bibnamefont{and}
  \bibinfo{author}{\bibfnamefont{B.}~\bibnamefont{Schenke}}
  (\bibinfo{year}{2011}{\natexlab{b}}), \eprint{1109.5992}.

\bibitem[{\citenamefont{Maldacena}(1998)}]{Maldacena:1997re}
\bibinfo{author}{\bibfnamefont{J.~M.} \bibnamefont{Maldacena}},
  \bibinfo{journal}{Adv. Theor. Math. Phys.} \textbf{\bibinfo{volume}{2}},
  \bibinfo{pages}{231} (\bibinfo{year}{1998}), \eprint{hep-th/9711200}.

\bibitem[{\citenamefont{Witten}(1998{\natexlab{a}})}]{Witten:1998qj}
\bibinfo{author}{\bibfnamefont{E.}~\bibnamefont{Witten}},
  \bibinfo{journal}{Adv. Theor. Math. Phys.} \textbf{\bibinfo{volume}{2}},
  \bibinfo{pages}{253} (\bibinfo{year}{1998}{\natexlab{a}}),
  \eprint{hep-th/9802150}.

\bibitem[{\citenamefont{Witten}(1998{\natexlab{b}})}]{Witten:1998zw}
\bibinfo{author}{\bibfnamefont{E.}~\bibnamefont{Witten}},
  \bibinfo{journal}{Adv. Theor. Math. Phys.} \textbf{\bibinfo{volume}{2}},
  \bibinfo{pages}{505} (\bibinfo{year}{1998}{\natexlab{b}}),
  \eprint{hep-th/9803131}.

\bibitem[{\citenamefont{Casalderrey-Solana
  et~al.}(2011{\natexlab{c}})\citenamefont{Casalderrey-Solana, Liu, Mateos,
  Rajagopal, and Wiedemann}}]{CasalderreySolana:2011us}
\bibinfo{author}{\bibfnamefont{J.}~\bibnamefont{Casalderrey-Solana}},
  \bibinfo{author}{\bibfnamefont{H.}~\bibnamefont{Liu}},
  \bibinfo{author}{\bibfnamefont{D.}~\bibnamefont{Mateos}},
  \bibinfo{author}{\bibfnamefont{K.}~\bibnamefont{Rajagopal}},
  \bibnamefont{and} \bibinfo{author}{\bibfnamefont{U.~A.}
  \bibnamefont{Wiedemann}} (\bibinfo{year}{2011}{\natexlab{c}}),
  \eprint{1101.0618}.

\bibitem[{\citenamefont{Hofman and Maldacena}(2008)}]{Hofman:2008ar}
\bibinfo{author}{\bibfnamefont{D.~M.} \bibnamefont{Hofman}} \bibnamefont{and}
  \bibinfo{author}{\bibfnamefont{J.}~\bibnamefont{Maldacena}},
  \bibinfo{journal}{JHEP} \textbf{\bibinfo{volume}{0805}}, \bibinfo{pages}{012}
  (\bibinfo{year}{2008}), \eprint{0803.1467}.

\bibitem[{\citenamefont{Gubser}(2006)}]{Gubser:2006bz}
\bibinfo{author}{\bibfnamefont{S.~S.} \bibnamefont{Gubser}},
  \bibinfo{journal}{Phys. Rev.} \textbf{\bibinfo{volume}{D74}},
  \bibinfo{pages}{126005} (\bibinfo{year}{2006}), \eprint{hep-th/0605182}.

\bibitem[{\citenamefont{Herzog et~al.}(2006)\citenamefont{Herzog, Karch,
  Kovtun, Kozcaz, and Yaffe}}]{Herzog:2006gh}
\bibinfo{author}{\bibfnamefont{C.}~\bibnamefont{Herzog}},
  \bibinfo{author}{\bibfnamefont{A.}~\bibnamefont{Karch}},
  \bibinfo{author}{\bibfnamefont{P.}~\bibnamefont{Kovtun}},
  \bibinfo{author}{\bibfnamefont{C.}~\bibnamefont{Kozcaz}}, \bibnamefont{and}
  \bibinfo{author}{\bibfnamefont{L.}~\bibnamefont{Yaffe}},
  \bibinfo{journal}{JHEP} \textbf{\bibinfo{volume}{0607}}, \bibinfo{pages}{013}
  (\bibinfo{year}{2006}), \eprint{hep-th/0605158}.

\bibitem[{\citenamefont{Casalderrey-Solana and
  Teaney}(2006)}]{CasalderreySolana:2006rq}
\bibinfo{author}{\bibfnamefont{J.}~\bibnamefont{Casalderrey-Solana}}
  \bibnamefont{and} \bibinfo{author}{\bibfnamefont{D.}~\bibnamefont{Teaney}},
  \bibinfo{journal}{Phys. Rev.} \textbf{\bibinfo{volume}{D74}},
  \bibinfo{pages}{085012} (\bibinfo{year}{2006}), \eprint{hep-ph/0605199}.

\bibitem[{\citenamefont{Gubser}(2008)}]{Gubser:2006nz}
\bibinfo{author}{\bibfnamefont{S.~S.} \bibnamefont{Gubser}},
  \bibinfo{journal}{Nucl. Phys.} \textbf{\bibinfo{volume}{B790}},
  \bibinfo{pages}{175} (\bibinfo{year}{2008}), \eprint{hep-th/0612143}.

\bibitem[{\citenamefont{Casalderrey-Solana and
  Teaney}(2007)}]{CasalderreySolana:2007qw}
\bibinfo{author}{\bibfnamefont{J.}~\bibnamefont{Casalderrey-Solana}}
  \bibnamefont{and} \bibinfo{author}{\bibfnamefont{D.}~\bibnamefont{Teaney}},
  \bibinfo{journal}{JHEP} \textbf{\bibinfo{volume}{0704}}, \bibinfo{pages}{039}
  (\bibinfo{year}{2007}), \eprint{hep-th/0701123}.

\bibitem[{\citenamefont{Friess et~al.}(2007)\citenamefont{Friess, Gubser,
  Michalogiorgakis, and Pufu}}]{Friess:2006fk}
\bibinfo{author}{\bibfnamefont{J.~J.} \bibnamefont{Friess}},
  \bibinfo{author}{\bibfnamefont{S.~S.} \bibnamefont{Gubser}},
  \bibinfo{author}{\bibfnamefont{G.}~\bibnamefont{Michalogiorgakis}},
  \bibnamefont{and} \bibinfo{author}{\bibfnamefont{S.~S.} \bibnamefont{Pufu}},
  \bibinfo{journal}{Phys. Rev.} \textbf{\bibinfo{volume}{D75}},
  \bibinfo{pages}{106003} (\bibinfo{year}{2007}), \eprint{hep-th/0607022}.

\bibitem[{\citenamefont{Yarom}(2007)}]{Yarom:2007ni}
\bibinfo{author}{\bibfnamefont{A.}~\bibnamefont{Yarom}},
  \bibinfo{journal}{Phys. Rev.} \textbf{\bibinfo{volume}{D75}},
  \bibinfo{pages}{105023} (\bibinfo{year}{2007}), \eprint{hep-th/0703095}.

\bibitem[{\citenamefont{Gubser et~al.}(2007)\citenamefont{Gubser, Pufu, and
  Yarom}}]{Gubser:2007xz}
\bibinfo{author}{\bibfnamefont{S.~S.} \bibnamefont{Gubser}},
  \bibinfo{author}{\bibfnamefont{S.~S.} \bibnamefont{Pufu}}, \bibnamefont{and}
  \bibinfo{author}{\bibfnamefont{A.}~\bibnamefont{Yarom}},
  \bibinfo{journal}{JHEP} \textbf{\bibinfo{volume}{0709}}, \bibinfo{pages}{108}
  (\bibinfo{year}{2007}), \eprint{0706.0213}.

\bibitem[{\citenamefont{Chesler and Yaffe}(2007)}]{Chesler:2007an}
\bibinfo{author}{\bibfnamefont{P.~M.} \bibnamefont{Chesler}} \bibnamefont{and}
  \bibinfo{author}{\bibfnamefont{L.~G.} \bibnamefont{Yaffe}},
  \bibinfo{journal}{Phys. Rev. Lett.} \textbf{\bibinfo{volume}{99}},
  \bibinfo{pages}{152001} (\bibinfo{year}{2007}), \eprint{0706.0368}.

\bibitem[{\citenamefont{Gubser et~al.}(2008{\natexlab{a}})\citenamefont{Gubser,
  Pufu, and Yarom}}]{Gubser:2007ga}
\bibinfo{author}{\bibfnamefont{S.~S.} \bibnamefont{Gubser}},
  \bibinfo{author}{\bibfnamefont{S.~S.} \bibnamefont{Pufu}}, \bibnamefont{and}
  \bibinfo{author}{\bibfnamefont{A.}~\bibnamefont{Yarom}},
  \bibinfo{journal}{Phys. Rev. Lett.} \textbf{\bibinfo{volume}{100}},
  \bibinfo{pages}{012301} (\bibinfo{year}{2008}{\natexlab{a}}),
  \eprint{0706.4307}.

\bibitem[{\citenamefont{Gubser and Yarom}(2008)}]{Gubser:2007ni}
\bibinfo{author}{\bibfnamefont{S.~S.} \bibnamefont{Gubser}} \bibnamefont{and}
  \bibinfo{author}{\bibfnamefont{A.}~\bibnamefont{Yarom}},
  \bibinfo{journal}{Phys. Rev.} \textbf{\bibinfo{volume}{D77}},
  \bibinfo{pages}{066007} (\bibinfo{year}{2008}), \eprint{0709.1089}.

\bibitem[{\citenamefont{Chesler and Yaffe}(2008)}]{Chesler:2007sv}
\bibinfo{author}{\bibfnamefont{P.~M.} \bibnamefont{Chesler}} \bibnamefont{and}
  \bibinfo{author}{\bibfnamefont{L.~G.} \bibnamefont{Yaffe}},
  \bibinfo{journal}{Phys. Rev.} \textbf{\bibinfo{volume}{D78}},
  \bibinfo{pages}{045013} (\bibinfo{year}{2008}), \eprint{0712.0050}.

\bibitem[{\citenamefont{Noronha et~al.}(2009)\citenamefont{Noronha, Gyulassy,
  and Torrieri}}]{Noronha:2008un}
\bibinfo{author}{\bibfnamefont{J.}~\bibnamefont{Noronha}},
  \bibinfo{author}{\bibfnamefont{M.}~\bibnamefont{Gyulassy}}, \bibnamefont{and}
  \bibinfo{author}{\bibfnamefont{G.}~\bibnamefont{Torrieri}},
  \bibinfo{journal}{Phys. Rev. Lett.} \textbf{\bibinfo{volume}{102}},
  \bibinfo{pages}{102301} (\bibinfo{year}{2009}), \eprint{0807.1038}.

\bibitem[{\citenamefont{Gubser et~al.}(2008{\natexlab{b}})\citenamefont{Gubser,
  Gulotta, Pufu, and Rocha}}]{Gubser:2008as}
\bibinfo{author}{\bibfnamefont{S.~S.} \bibnamefont{Gubser}},
  \bibinfo{author}{\bibfnamefont{D.~R.} \bibnamefont{Gulotta}},
  \bibinfo{author}{\bibfnamefont{S.~S.} \bibnamefont{Pufu}}, \bibnamefont{and}
  \bibinfo{author}{\bibfnamefont{F.~D.} \bibnamefont{Rocha}},
  \bibinfo{journal}{JHEP} \textbf{\bibinfo{volume}{0810}}, \bibinfo{pages}{052}
  (\bibinfo{year}{2008}{\natexlab{b}}), \eprint{0803.1470}.

\bibitem[{\citenamefont{Chesler
  et~al.}(2009{\natexlab{a}})\citenamefont{Chesler, Jensen, and
  Karch}}]{Chesler:2008wd}
\bibinfo{author}{\bibfnamefont{P.~M.} \bibnamefont{Chesler}},
  \bibinfo{author}{\bibfnamefont{K.}~\bibnamefont{Jensen}}, \bibnamefont{and}
  \bibinfo{author}{\bibfnamefont{A.}~\bibnamefont{Karch}},
  \bibinfo{journal}{Phys. Rev.} \textbf{\bibinfo{volume}{D79}},
  \bibinfo{pages}{025021} (\bibinfo{year}{2009}{\natexlab{a}}),
  \eprint{0804.3110}.

\bibitem[{\citenamefont{Chesler
  et~al.}(2009{\natexlab{b}})\citenamefont{Chesler, Jensen, Karch, and
  Yaffe}}]{Chesler:2008uy}
\bibinfo{author}{\bibfnamefont{P.~M.} \bibnamefont{Chesler}},
  \bibinfo{author}{\bibfnamefont{K.}~\bibnamefont{Jensen}},
  \bibinfo{author}{\bibfnamefont{A.}~\bibnamefont{Karch}}, \bibnamefont{and}
  \bibinfo{author}{\bibfnamefont{L.~G.} \bibnamefont{Yaffe}},
  \bibinfo{journal}{Phys. Rev.} \textbf{\bibinfo{volume}{D79}},
  \bibinfo{pages}{125015} (\bibinfo{year}{2009}{\natexlab{b}}),
  \eprint{0810.1985}.

\bibitem[{\citenamefont{Liu et~al.}(2006)\citenamefont{Liu, Rajagopal, and
  Wiedemann}}]{Liu:2006ug}
\bibinfo{author}{\bibfnamefont{H.}~\bibnamefont{Liu}},
  \bibinfo{author}{\bibfnamefont{K.}~\bibnamefont{Rajagopal}},
  \bibnamefont{and} \bibinfo{author}{\bibfnamefont{U.~A.}
  \bibnamefont{Wiedemann}}, \bibinfo{journal}{Phys. Rev. Lett.}
  \textbf{\bibinfo{volume}{97}}, \bibinfo{pages}{182301}
  (\bibinfo{year}{2006}), \eprint{hep-ph/0605178}.

\bibitem[{\citenamefont{D'Eramo et~al.}(2011)\citenamefont{D'Eramo, Liu, and
  Rajagopal}}]{D'Eramo:2010ak}
\bibinfo{author}{\bibfnamefont{F.}~\bibnamefont{D'Eramo}},
  \bibinfo{author}{\bibfnamefont{H.}~\bibnamefont{Liu}}, \bibnamefont{and}
  \bibinfo{author}{\bibfnamefont{K.}~\bibnamefont{Rajagopal}},
  \bibinfo{journal}{Phys. Rev.} \textbf{\bibinfo{volume}{D84}},
  \bibinfo{pages}{065015} (\bibinfo{year}{2011}), \eprint{1006.1367}.

\bibitem[{\citenamefont{Athanasiou et~al.}(2010)\citenamefont{Athanasiou,
  Chesler, Liu, Nickel, and Rajagopal}}]{Athanasiou:2010pv}
\bibinfo{author}{\bibfnamefont{C.}~\bibnamefont{Athanasiou}},
  \bibinfo{author}{\bibfnamefont{P.~M.} \bibnamefont{Chesler}},
  \bibinfo{author}{\bibfnamefont{H.}~\bibnamefont{Liu}},
  \bibinfo{author}{\bibfnamefont{D.}~\bibnamefont{Nickel}}, \bibnamefont{and}
  \bibinfo{author}{\bibfnamefont{K.}~\bibnamefont{Rajagopal}},
  \bibinfo{journal}{Phys. Rev.} \textbf{\bibinfo{volume}{D81}},
  \bibinfo{pages}{126001} (\bibinfo{year}{2010}), \eprint{1001.3880}.

\bibitem[{\citenamefont{Hubeny}(2011)}]{Hubeny:2010bq}
\bibinfo{author}{\bibfnamefont{V.~E.} \bibnamefont{Hubeny}},
  \bibinfo{journal}{New J.Phys.} \textbf{\bibinfo{volume}{13}},
  \bibinfo{pages}{035006} (\bibinfo{year}{2011}), \eprint{1012.3561}.

\bibitem[{\citenamefont{Hatta et~al.}(2011{\natexlab{a}})\citenamefont{Hatta,
  Iancu, Mueller, and Triantafyllopoulos}}]{Hatta:2010dz}
\bibinfo{author}{\bibfnamefont{Y.}~\bibnamefont{Hatta}},
  \bibinfo{author}{\bibfnamefont{E.}~\bibnamefont{Iancu}},
  \bibinfo{author}{\bibfnamefont{A.}~\bibnamefont{Mueller}}, \bibnamefont{and}
  \bibinfo{author}{\bibfnamefont{D.}~\bibnamefont{Triantafyllopoulos}},
  \bibinfo{journal}{JHEP} \textbf{\bibinfo{volume}{1102}}, \bibinfo{pages}{065}
  (\bibinfo{year}{2011}{\natexlab{a}}), \eprint{1011.3763}.

\bibitem[{\citenamefont{Hatta et~al.}(2011{\natexlab{b}})\citenamefont{Hatta,
  Iancu, Mueller, and Triantafyllopoulos}}]{Hatta:2011gh}
\bibinfo{author}{\bibfnamefont{Y.}~\bibnamefont{Hatta}},
  \bibinfo{author}{\bibfnamefont{E.}~\bibnamefont{Iancu}},
  \bibinfo{author}{\bibfnamefont{A.}~\bibnamefont{Mueller}}, \bibnamefont{and}
  \bibinfo{author}{\bibfnamefont{D.}~\bibnamefont{Triantafyllopoulos}},
  \bibinfo{journal}{Nucl. Phys.} \textbf{\bibinfo{volume}{B850}},
  \bibinfo{pages}{31} (\bibinfo{year}{2011}{\natexlab{b}}), \eprint{1102.0232}.

\bibitem[{\citenamefont{Chernicoff
  et~al.}(2011{\natexlab{a}})\citenamefont{Chernicoff, Guijosa, and
  Pedraza}}]{Chernicoff:2011vn}
\bibinfo{author}{\bibfnamefont{M.}~\bibnamefont{Chernicoff}},
  \bibinfo{author}{\bibfnamefont{A.}~\bibnamefont{Guijosa}}, \bibnamefont{and}
  \bibinfo{author}{\bibfnamefont{J.~F.} \bibnamefont{Pedraza}}
  (\bibinfo{year}{2011}{\natexlab{a}}), \eprint{1106.4059}.

\bibitem[{\citenamefont{Baier}(2011)}]{Baier:2011dh}
\bibinfo{author}{\bibfnamefont{R.}~\bibnamefont{Baier}} (\bibinfo{year}{2011}),
  \eprint{1107.4250}.

\bibitem[{\citenamefont{Chernicoff
  et~al.}(2011{\natexlab{b}})\citenamefont{Chernicoff, Garcia, Guijosa, and
  Pedraza}}]{Chernicoff:2011xv}
\bibinfo{author}{\bibfnamefont{M.}~\bibnamefont{Chernicoff}},
  \bibinfo{author}{\bibfnamefont{J.}~\bibnamefont{Garcia}},
  \bibinfo{author}{\bibfnamefont{A.}~\bibnamefont{Guijosa}}, \bibnamefont{and}
  \bibinfo{author}{\bibfnamefont{J.~F.} \bibnamefont{Pedraza}}
  (\bibinfo{year}{2011}{\natexlab{b}}), \eprint{1111.0872}.

\bibitem[{\citenamefont{Fadafan et~al.}(2009)\citenamefont{Fadafan, Liu,
  Rajagopal, and Wiedemann}}]{Fadafan:2008bq}
\bibinfo{author}{\bibfnamefont{K.~B.} \bibnamefont{Fadafan}},
  \bibinfo{author}{\bibfnamefont{H.}~\bibnamefont{Liu}},
  \bibinfo{author}{\bibfnamefont{K.}~\bibnamefont{Rajagopal}},
  \bibnamefont{and} \bibinfo{author}{\bibfnamefont{U.~A.}
  \bibnamefont{Wiedemann}}, \bibinfo{journal}{Eur. Phys. J.}
  \textbf{\bibinfo{volume}{C61}}, \bibinfo{pages}{553} (\bibinfo{year}{2009}),
  \eprint{0809.2869}.

\bibitem[{\citenamefont{Mikhailov}(2003)}]{Mikhailov:2003er}
\bibinfo{author}{\bibfnamefont{A.}~\bibnamefont{Mikhailov}}
  (\bibinfo{year}{2003}), \eprint{hep-th/0305196}.

\bibitem[{\citenamefont{Connors}(2011)}]{Connors:2011zz}
\bibinfo{author}{\bibfnamefont{M.}~\bibnamefont{Connors}}
  (\bibinfo{collaboration}{PHENIX Collaboration}), \bibinfo{journal}{Nucl.
  Phys.} \textbf{\bibinfo{volume}{A855}}, \bibinfo{pages}{335}
  (\bibinfo{year}{2011}).

\bibitem[{\citenamefont{Caines}(2011{\natexlab{a}})}]{Caines:2011xp}
\bibinfo{author}{\bibfnamefont{H.}~\bibnamefont{Caines}}
  (\bibinfo{collaboration}{STAR Collaboration})
  (\bibinfo{year}{2011}{\natexlab{a}}), \eprint{1106.6247}.

\bibitem[{\citenamefont{Purschke}(2011)}]{Purschke:2011xx}
\bibinfo{author}{\bibfnamefont{M.}~\bibnamefont{Purschke}}
  (\bibinfo{collaboration}{PHENIX collaboration}) (\bibinfo{year}{2011}),
  \eprint{to appear in QM2011 proceedings}.

\bibitem[{\citenamefont{Ohlson}(2011)}]{Ohlson:2011xn}
\bibinfo{author}{\bibfnamefont{A.}~\bibnamefont{Ohlson}}
  (\bibinfo{collaboration}{STAR Collaboration}) (\bibinfo{year}{2011}),
  \eprint{1106.6243}.

\bibitem[{\citenamefont{Caines}(2011{\natexlab{b}})}]{Caines:2011ew}
\bibinfo{author}{\bibfnamefont{H.}~\bibnamefont{Caines}}
  (\bibinfo{year}{2011}{\natexlab{b}}), \eprint{1110.1878}.

\bibitem[{\citenamefont{Boyd}(2001)}]{Boyd:2001}
\bibinfo{author}{\bibfnamefont{J.~P.} \bibnamefont{Boyd}},
  \emph{\bibinfo{title}{Chebyshev and Fourier spectral methods}}
  (\bibinfo{publisher}{{Dover}}, \bibinfo{year}{2001}), \bibinfo{edition}{2nd}
  ed.

\bibitem[{\citenamefont{de~Haro et~al.}(2001)\citenamefont{de~Haro, Solodukhin,
  and Skenderis}}]{deHaro:2000xn}
\bibinfo{author}{\bibfnamefont{S.}~\bibnamefont{de~Haro}},
  \bibinfo{author}{\bibfnamefont{S.~N.} \bibnamefont{Solodukhin}},
  \bibnamefont{and}
  \bibinfo{author}{\bibfnamefont{K.}~\bibnamefont{Skenderis}},
  \bibinfo{journal}{Commun. Math. Phys.} \textbf{\bibinfo{volume}{217}},
  \bibinfo{pages}{595} (\bibinfo{year}{2001}), \eprint{hep-th/0002230}.

\bibitem[{\citenamefont{Baier et~al.}(2008)\citenamefont{Baier, Romatschke,
  Son, Starinets, and Stephanov}}]{Baier:2007ix}
\bibinfo{author}{\bibfnamefont{R.}~\bibnamefont{Baier}},
  \bibinfo{author}{\bibfnamefont{P.}~\bibnamefont{Romatschke}},
  \bibinfo{author}{\bibfnamefont{D.~T.} \bibnamefont{Son}},
  \bibinfo{author}{\bibfnamefont{A.~O.} \bibnamefont{Starinets}},
  \bibnamefont{and} \bibinfo{author}{\bibfnamefont{M.~A.}
  \bibnamefont{Stephanov}}, \bibinfo{journal}{JHEP}
  \textbf{\bibinfo{volume}{04}}, \bibinfo{pages}{100} (\bibinfo{year}{2008}),
  \eprint{0712.2451}.

\bibitem[{\citenamefont{Kovtun and Starinets}(2005)}]{Kovtun:2005ev}
\bibinfo{author}{\bibfnamefont{P.~K.} \bibnamefont{Kovtun}} \bibnamefont{and}
  \bibinfo{author}{\bibfnamefont{A.~O.} \bibnamefont{Starinets}},
  \bibinfo{journal}{Phys. Rev.} \textbf{\bibinfo{volume}{D72}},
  \bibinfo{pages}{086009} (\bibinfo{year}{2005}), \eprint{hep-th/0506184}.

\bibitem[{\citenamefont{Festuccia and Liu}(2008)}]{Festuccia:2008zx}
\bibinfo{author}{\bibfnamefont{G.}~\bibnamefont{Festuccia}} \bibnamefont{and}
  \bibinfo{author}{\bibfnamefont{H.}~\bibnamefont{Liu}} (\bibinfo{year}{2008}),
  \eprint{0811.1033}.

\bibitem[{\citenamefont{Horowitz and Gyulassy}(2011)}]{Horowitz:2011cv}
\bibinfo{author}{\bibfnamefont{W.}~\bibnamefont{Horowitz}} \bibnamefont{and}
  \bibinfo{author}{\bibfnamefont{M.}~\bibnamefont{Gyulassy}}
  (\bibinfo{year}{2011}), \eprint{1107.2136}.

\bibitem[{\citenamefont{Bathe}(2011)}]{Bathe:2011xx}
\bibinfo{author}{\bibfnamefont{S.}~\bibnamefont{Bathe}}
  (\bibinfo{collaboration}{PHENIX collaboration}) (\bibinfo{year}{2011}),
  \eprint{to appear in QM2011 proceedings}.

\bibitem[{\citenamefont{Chernicoff and Guijosa}(2008)}]{Chernicoff:2008sa}
\bibinfo{author}{\bibfnamefont{M.}~\bibnamefont{Chernicoff}} \bibnamefont{and}
  \bibinfo{author}{\bibfnamefont{A.}~\bibnamefont{Guijosa}},
  \bibinfo{journal}{JHEP} \textbf{\bibinfo{volume}{0806}}, \bibinfo{pages}{005}
  (\bibinfo{year}{2008}), \eprint{0803.3070}.

\bibitem[{\citenamefont{Dominguez et~al.}(2008)\citenamefont{Dominguez,
  Marquet, Mueller, Wu, and Xiao}}]{Dominguez:2008vd}
\bibinfo{author}{\bibfnamefont{F.}~\bibnamefont{Dominguez}},
  \bibinfo{author}{\bibfnamefont{C.}~\bibnamefont{Marquet}},
  \bibinfo{author}{\bibfnamefont{A.}~\bibnamefont{Mueller}},
  \bibinfo{author}{\bibfnamefont{B.}~\bibnamefont{Wu}}, \bibnamefont{and}
  \bibinfo{author}{\bibfnamefont{B.-W.} \bibnamefont{Xiao}},
  \bibinfo{journal}{Nucl.Phys.} \textbf{\bibinfo{volume}{A811}},
  \bibinfo{pages}{197} (\bibinfo{year}{2008}), \eprint{0803.3234}.

\bibitem[{\citenamefont{Xiao}(2008)}]{Xiao:2008nr}
\bibinfo{author}{\bibfnamefont{B.-W.} \bibnamefont{Xiao}},
  \bibinfo{journal}{Phys.Lett.} \textbf{\bibinfo{volume}{B665}},
  \bibinfo{pages}{173} (\bibinfo{year}{2008}), \eprint{0804.1343}.

\bibitem[{\citenamefont{Beuf et~al.}(2009)\citenamefont{Beuf, Marquet, and
  Xiao}}]{Beuf:2008ep}
\bibinfo{author}{\bibfnamefont{G.}~\bibnamefont{Beuf}},
  \bibinfo{author}{\bibfnamefont{C.}~\bibnamefont{Marquet}}, \bibnamefont{and}
  \bibinfo{author}{\bibfnamefont{B.-W.} \bibnamefont{Xiao}},
  \bibinfo{journal}{Phys.Rev.} \textbf{\bibinfo{volume}{D80}},
  \bibinfo{pages}{085001} (\bibinfo{year}{2009}), \eprint{0812.1051}.

\end{thebibliography}

\end{document}